\documentclass[a4paper,11pt]{article}
\pdfoutput=1 

\usepackage{jheppub} 

\usepackage[T1]{fontenc} 

\newcommand*{\ddd}{$\Delta \delta_D$}
\newcommand*{\kl}{$K_{\rm L}^0$}
\newcommand*{\ks}{$K_{\rm S}^0$}
\newcommand*{\klpipi}{$K_{\rm L}^0\pi^+\pi^-$}
\newcommand*{\kspipi}{$K_{\rm S}^0\pi^+\pi^-$}
\newcommand*{\kslpipi}{$K_{\rm S(L)}^0\pi^+\pi^-$}

\newcommand*{\ksorlpipi}{$K_{\rm S/L}^0\pi^+\pi^-$}
\newcommand*{\dtoklpipi}{$D^0 \to K_{\rm L}^0\pi^+\pi^-$}
\newcommand*{\dtokspipi}{$D^0 \to K_{\rm S}^0\pi^+\pi^-$}
\newcommand*{\dbartoklpipi}{$\bar{D}^0 \to K_{\rm L}^0\pi^+\pi^-$}

\newcommand*{\dtokslpipi}{$D^0 \to K_{\rm S,L}^0\pi^+\pi^-$}
\newcommand*{\cicip}{$c_i^{(\mathcal{0})}$}
\newcommand*{\sisip}{$s_i^{(\mathcal{0})}$}

\newcommand\Tstrut{\rule{0pt}{3.5ex}}         
\newcommand\Bstrut{\rule[-2.0ex]{0pt}{0pt}}   

\newcommand\Tstrutsmall{\rule{0pt}{2.2ex}}         
\newcommand\Bstrutsmall{\rule[-1.0ex]{0pt}{0pt}}   

\usepackage{mathtools}

\DeclarePairedDelimiter\ket{\lvert}{\rangle}
\DeclarePairedDelimiterX\braket[2]{\langle}{\rangle}{#1 \delimsize\vert #2}

\usepackage{caption}
\usepackage{subcaption}

\usepackage{lineno}

\title{\boldmath Determination of U-spin breaking parameters with an amplitude analysis of the decay $D^0 \to K_{\rm L}^0\pi^+\pi^-$}

\author{The BESIII collaboration}

\emailAdd{besiii-publications@ihep.ac.cn}

\abstract{We present a study of the resonant structure of the decay $D^0 \to K_{\rm L}^0\pi^+\pi^-$, using quantum-correlated $D^0\bar{D}^0$ data produced at $\sqrt{s}=3.773$ GeV. The data sample was collected by the BESIII experiment and corresponds to an integrated luminosity of $2.93$~fb$^{-1}$. This study is the first amplitude analysis of a decay mode involving a $K_{\rm L}^0$, which also results in the first measurement of the complex {\it U-spin breaking parameters} ($\hat{\rho}$) related to various $\mathit{CP}$-eigenstate resonant modes through which the three-body decay proceeds. The moduli of the $\hat{\rho}$ parameters have central values in a wide range from $0.4$ to $12.1$, which indicates substantial U-spin symmetry breaking. We present the fractional resonant contributions and average strong-phase parameters over regions of phase space for both $K_{\rm S}^0\pi^+\pi^-$ and $K_{\rm L}^0\pi^+\pi^-$ modes. We also report the ratio of the branching fractions between \klpipi\ and \kspipi\ decay modes and the $\mathit{CP}$-even fraction of the \klpipi\ state calculated using the U-spin breaking parameters.}

\let\oldequation\equation
\let\oldendequation\endequation
\renewenvironment{equation}
  {\linenomathNonumbers\oldequation}
  {\oldendequation\endlinenomath}

\begin{document} 

\maketitle

\clearpage

\begin{center}
M.~Ablikim$^{1}$, M.~N.~Achasov$^{12,b}$, P.~Adlarson$^{72}$, R.~Aliberti$^{33}$, A.~Amoroso$^{71A,71C}$, M.~R.~An$^{37}$, Q.~An$^{68,55}$, Y.~Bai$^{54}$, O.~Bakina$^{34}$, R.~Baldini Ferroli$^{27A}$
% ...
\\
\vspace{0.2cm}
(BESIII Collaboration)\\
\vspace{0.2cm} {\it
$^{1}$ Institute of High Energy Physics, Beijing 100049, People's Republic of China\\
% ...
$^{12}$ G.I. Budker Institute of Nuclear Physics SB RAS (BINP), Novosibirsk 630090, Russia\\
% ...
$^{27}$ INFN Laboratori Nazionali di Frascati , (A)INFN Laboratori Nazionali di Frascati, I-00044, Frascati, Italy; (B)INFN Sezione di  Perugia, I-06100, Perugia, Italy; (C)University of Perugia, I-06100, Perugia, Italy\\
% ...
$^{33}$ Johannes Gutenberg University of Mainz, Johann-Joachim-Becher-Weg 45, D-55099 Mainz, Germany\\
$^{34}$ Joint Institute for Nuclear Research, 141980 Dubna, Moscow region, Russia\\
% ...
$^{37}$ Liaoning Normal University, Dalian 116029, People's Republic of China\\
% ...
$^{54}$ Southeast University, Nanjing 211100, People's Republic of China\\
$^{55}$ State Key Laboratory of Particle Detection and Electronics, Beijing 100049, Hefei 230026, People's Republic of China\\
% ...
$^{68}$ University of Science and Technology of China, Hefei 230026, People's Republic of China\\
% ...
$^{71}$ University of Turin and INFN, (A)University of Turin, I-10125, Turin, Italy; (B)University of Eastern Piedmont, I-15121, Alessandria, Italy; (C)INFN, I-10125, Turin, Italy\\
$^{72}$ Uppsala University, Box 516, SE-75120 Uppsala, Sweden\\
% ...
$^{b}$ Also at the Novosibirsk State University, Novosibirsk, 630090, Russia\\
% ...

}

\end{center}

\clearpage

\flushbottom

\section{Introduction}
\label{sec:intro}

The phenomenon of $\mathit{CP}$ violation in the standard model (SM) is parametrized by a single irreducible phase in the complex Cabibbo-Kobayashi-Maskawa (CKM) quark-mixing matrix~\cite{ref:ckm1, ref:ckm2}, which describes the weak interaction of quarks. Exploiting the unitary nature of the CKM matrix, this $\mathit{CP}$-violating phase can be represented on the complex plane as the argument of a particular combination of the CKM elements $V_{qq^{\prime}}$. The phase  $-\arg\left[(V_{ub}^{*}V_{us})/(V_{cb}^{*}V_{cs})\right]$ is denoted by $\gamma$ and can be measured by studying interference between decays with identical final states where one proceeds via a $b \to u$ transition~\cite{ref:glw, ref:atwoodsoni97, ref:atwoodsoni01}. The decay $B^{\pm}\to Dh^{\pm}$, where $h^{\pm}$ denotes a $K^{\pm}$ or $\pi^{\pm}$ and $D$ a superposition of the flavor states of neutral $D$ meson, proceeds almost purely at tree-level; electroweak box and loop corrections are below $\mathcal{O}(10^{-7})$~\cite{ref:GamThErr}, thereby excluding the possibility of loop-level contributions from beyond-the-SM physics~\cite{ref:NPintree}. The absence of theoretical uncertainties makes this channel ideal to determine $\gamma$. The $\gamma$ determination method put forward in Refs.~\cite{ref:bpggsz, ref:bondar_proc} requires the $D$ meson to decay into self-conjugate multi-body final states such as \kspipi. Such multi-body $D$-meson decays provide regions of phase space where interference between $CP$-eigenstates of the $D$ meson~\cite{ref:glw}, and Cabibbo-favoured (CF) and doubly Cabibbo-suppressed (DCS)~\cite{ref:atwoodsoni97} decays take place, which allows $\gamma$ and the strong-dynamics of the $B$ decay to be extracted from a single decay mode. 

Although the current world average of $\gamma$ is still statistically limited, the statistical uncertainty has reduced by approximately a factor eight over the last decade, and the ultimate data samples of LHCb and Belle II should result in a statistical uncertainty in $\gamma$ close to $1^{\circ}$. The primary source of systematic uncertainty is inputs from the $D$-decay parameters \cite{ref:Gamlhcb21, ref:Gambelle22}. These inputs are the strong-phase differences \ddd, between $D^0$ and $\bar{D}^0$ decays which are measured in quantum-correlated $D$ decay~\cite{ref:bondar, ref:AtwoodSoniCharm}. 
The strong-phase difference arising from the interference of $D^0$ and $\bar{D}^0$ decaying into a common final state \kspipi\ is a critical input not only to the $\gamma$ measurement using $B^{\pm} \to DK^{\pm}$ decay channels but also to other important flavor studies: the time-dependent measurement of the CKM angle $\beta$ through $B^0 \to \bar{D}^{(*)0}h^0$ decays~\cite{ref:phi1belle, ref:bellebabar18} and the measurement of $\mathit{CP}$ violation and mixing in neutral $D$ meson system~\cite{ref:Dmixinglhcb}.

Quantum-correlated $D\bar{D}$ events at $\psi(3770)$ recorded at BESIII give access to the strong-phase difference when $D$ decays are reconstructed by means of flavor tagging~\cite{ref:bes3exp}. The pairs of $D$ mesons are quantum-correlated because they are produced in a $J^{PC} = 1^{--}$ state with an anti-symmetric wavefunction,
\begin{equation}
  \ket{\psi(3770)} = \ket{D\bar{D}} = \frac{1}{\sqrt{2}}\left( \ket{D^0}\ket{\bar{D}^0} - \ket{\bar{D}^0} \ket{D^0} \right),
\end{equation}
 which constrains the decay product of one $D$ meson given the other, discussed more specifically in section~\ref{sec:theory}. A model-independent BESIII analysis measured the average sine and cosine of \ddd\ for \kspipi\ $\left(c_i,s_i\right)$ and \klpipi\ $\left(c_i^{\prime},s_i^{\prime}\right)$~\cite{ref:bes3prl_cisi,ref:bes3prd_cisi}. Inclusion of the \klpipi\ mode provides a three-times-larger data sample at BESIII due to higher \kl\ reconstruction efficiency and combinatorics of $D\bar{D} \to (K_{\rm S}^0\pi^+\pi^-)^2$ versus $D\bar{D} \to (K_{\rm S}^0\pi^+\pi^-, K_{\rm L}^0\pi^+\pi^-)$ decays. However, including these \klpipi decays introduces a systematic uncertainty related to assumptions about the values of complex {\it U-spin breaking} parameters ($\hat{\rho}$) that separate the decay amplitudes of \dtoklpipi\ and \dtokspipi\ modes. In previous analyses~\cite{ref:bes3prl_cisi,ref:bes3prd_cisi,ref:cleo_cisi}, the nominal value of the parameters was unity; and a systematic uncertainty on this assumption was derived by assuming  $\left|\hat{\rho}\right|$ had an uncertainty of $50\%$ and $\arg\left(\hat{\rho}\right)$ could have any value in the interval ($-180^{\circ}, 180^{\circ}$). These U-spin breaking parameters, which are discussed in greater detail in section~\ref{sec:theory}, have never been experimentally determined and the only way to measure them is through an amplitude analysis of the \dtoklpipi\ decay. This paper presents the first experimental measurements of $\hat{\rho}$ in this decay.

The remainder of this paper is organized as follows. A brief discussion about the model-independent measurement of strong-phase parameters and a review of amplitude parametrizations for a three-body decay are given in section~\ref{sec:theory}. An overview of the BESIII detector and the simulations performed for this analysis is given in section~\ref{sec:detdata}, while section~\ref{sec:evtsel} lists various event selection criteria adopted to select the data samples. The amplitude analysis implementation and validation are presented in section~\ref{sec:ampfit}. Results are given in section~\ref{sec:fitres}, while section~\ref{ssec:syst} presents a study of systematic uncertainties. Section~\ref{sec:cpcontent} provides additional model predictions in the form of $CP$ content of \kspipi\ and \klpipi\ modes and the ratio of their branching fractions in $D$ decays. Section~\ref{sec:conc} reports the conclusions.

\section{Strong-phase and amplitude-analysis formalism}
\label{sec:theory}
 In this section we first define the strong-phase parameters and how those for $D\to K_{\rm S}^0\pi^+\pi^-$ and $D\to K_{\rm L}^0\pi^+\pi^-$ are related together. Then we discuss the amplitude-analysis technique employed to determine the $\hat{\rho}$ parameters.
 
\begin{figure}[t!]
    \centering
    \includegraphics[scale=0.3]{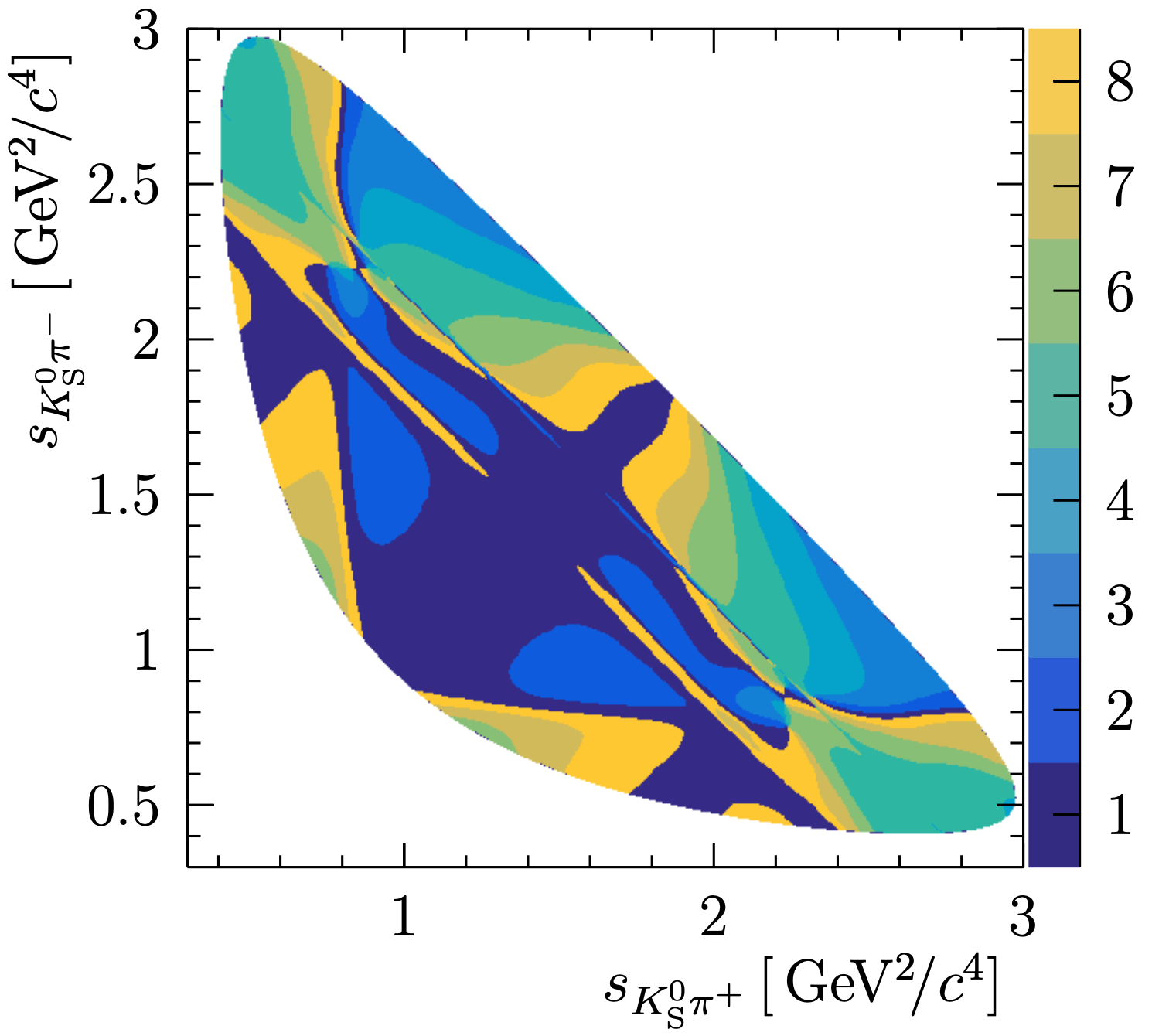}
    \caption{Equal-$\Delta\delta_D$ bins on the Dalitz plot of $D^0 \to K_{\rm S}^0\pi^+\pi^-$ decay.}
    \label{fig:DPbins}
\end{figure}

The $D$-decay parameters that appear in these studies are cosines and sines of the strong-phase difference averaged over regions of phase space. The two-dimensional phase space of \kspipi\ and \klpipi\ decay modes is described in terms of pairwise invariant masses of the final-state particles. Of the three possible permutations, only two will be independent, forming a Dalitz plot (DP) for which the phase space is uniform within its boundaries. In this paper we use invariant squared masses of $K_{\rm S(L)}^0\pi^+$ and $K_{\rm S(L)}^0\pi^-$, which are written as $s_{K_{\rm S(L)}^0\pi^+}$ and $s_{K_{\rm S(L)}^0\pi^-}$, respectively. The DP is divided into bins to gain sensitivity to the large variations in the strong-phase difference $\Delta\delta_D$ across the DP. One common binning scheme is the {\it equal-\ddd} binning that minimizes the variation in the values of $\Delta\delta_D$ in each bin; this scheme is shown in figure~\ref{fig:DPbins}. The DP is divided into sixteen bins, which are symmetric about the line $s_{K_{\rm S(L)}^0\pi^+}=s_{K_{\rm S(L)}^0\pi^-}$.
The weighted averages of cosines and sines of the strong-phase difference in the $i^{\rm th}$ bin of the DP are given by
\begin{equation}
c_i = \frac{1}{\sqrt{\int_i |A_D(\textbf{x})|^2 {\rm d}\textbf{x}~ \int_i |A_{\bar{D}}(\textbf{x})|^2 {\rm d}\textbf{x} }}~ \int_i |A_D({\textbf{x}})| |A_{\bar{D}}({\textbf{x}})| {\rm cos}(\Delta \delta_D) {\rm d}\textbf{x},
\label{eq:cisidef}
\end{equation}
and an analogous expression with sine of the strong-phase difference, where $A_D(\textbf{x})$ and $A_{\bar{D}}(\textbf{x})$ are the decay amplitudes of $D^0$ and $\bar{D}^0$, respectively, at point $\mathbf{x}=\left(s_{K_{\rm S}^0\pi^+},\:s_{K_{\rm S}^0\pi^-}\right)$ in the same bin on the DP of the final state \kspipi. A similar definition of strong-phase parameters $c_i'$ and $s_i'$ can be written for the \klpipi\ decay mode. The primed parameters henceforth correspond to the \klpipi\ mode. 

In a model-independent measurement of \cicip\ and \sisip\ with quantum-correlated $D\bar{D}$ events, the observables are yields of events for which the decays of both the $D$ meson states are reconstructed, known as {\it double-tagged} (DT) yields. More precisely, the first set of observables are the expected yields of \kspipi\ or \klpipi\ in  the $i^{\rm th}$ DP bin that are reconstructed against an exact or approximate $\mathit{CP}$ eigenstate such as $K^+K^-$ or $\pi^+\pi^-\pi^0$. These observables, conventionally denoted by $M_i^{(\mathcal{0})}$ for the \kslpipi\ mode, are only sensitive to \cicip\ but not \sisip. The second set of observables are yields of the signal \kslpipi\ mode in the $i^{\rm th}$ DP bin reconstructed against another \kspipi\ mode in the $j^{\rm th}$ bin of its DP. These are denoted by $M_{ij}^{(\mathcal{0})}$ and are sensitive to both \cicip\ and \sisip. Furthermore, the measured parameter differences between \kspipi\ and \klpipi\ modes, $\Delta c_i^{\rm meas} = (c_i - c_i')$ and $\Delta s_i^{\rm meas} = (s_i - s_i')$ are constrained to their model-predicted values $\Delta c_i^{\rm pred}$ and $\Delta s_i^{\rm pred}$, respectively. The constraint is implemented via a $\chi^2$ penalty term:
\begin{equation}
\chi^2 = \sum_{i} \left[ \left( \Delta c_i^{\rm meas} - \Delta c_i^{\rm pred} \right)/ \delta(\Delta c_i^{\rm pred}) \right]^2 + \sum_i \left[ \left( \Delta s_i^{\rm meas} - \Delta s_i^{\rm pred} \right)/ \delta(\Delta s_i^{\rm pred}) \right]^2,
\end{equation}
where $\delta(\Delta c_i^{\rm pred})$ and $\delta (\Delta s_i^{\rm pred})$ are the associated uncertainties on the model-predicted differences. Including the $K_{\rm L}^0\pi^+\pi^-$ final state improves sensitivity, particularly to \sisip. 

The uncertainties on the model-predicted values of the differences, $\delta(\Delta c_i^{\rm pred})$ and $\delta (\Delta s_i^{\rm pred})$ are dominated by assumptions associated with the U-spin breaking parameters~\cite{ref:bes3prl_cisi, ref:bes3prd_cisi}. This uncertainty motivates an amplitude analysis of \dtoklpipi, which will test these assumptions and determine a well defined data-driven uncertainty. 

With the motivation for an amplitude analysis of $D^0\to K^0_{\rm L}\pi^+\pi^-$ described, we now provide the formalism for such an analysis. Any three-body decay $D\to abc$ can proceed via multiple quasi-independent two-body intermediate channels:
\begin{equation}
\label{eq:decay_chain}
    D \to rc,~ r \to ab,
\end{equation}
where $r$ is an intermediate resonance.
The total effective amplitude of this decay topology is given by a coherent sum of all the contributing resonant channels. This approximation is called the {\it isobar model}, where the contributing intermediate amplitudes are referred to as the {\it isobars}. Isobars can be modeled with various complex dynamical functions, the choice of which depends on the spin and width of the resonance. In addition to the resonant modes, the total decay amplitude may also include a three-body {\it non-resonant} channel:
\begin{equation}
\mathcal{A}(\textbf{x}) = \mathcal{A}_{\rm resonant} +  \mathcal{A}_{\rm NR} = \sum_r a_r e^{i\phi_r}A_r(\textbf{x}) + a_0e^{i\phi_0}, 
\end{equation}
where $\mathcal{A}(\textbf{x})$ is the final decay amplitude at position $\textbf{x}$ in the DP. Here the complex coupling parameters $a_re^{i\phi_r}$ correspond to resonant contributions denoted by $r$ and provide relative magnitudes and phases to each of these resonant amplitudes. As the DP phase space is uniform,  only the dynamical part $A_r(\textbf{x})$ of the total decay rate results in variations of event density over the DP. Nominally, the dynamics of the modes associated with well isolated and narrow resonant structures with spin one or two are described by relativistic {\it Breit-Wigner} functions. In contrast, dynamics of broad overlapping resonant structures, which usually is the case with scalars, are parametrized using the K-matrix formulation borrowed from scattering theory. For the subsequent discussions on various parametrizations in the rest of this section, a generic decay chain will be referred to, as in eq.~\ref{eq:decay_chain}, with an angular-momentum transfer $J_{D} \to j_{r} + L$, where $J_D$ and $j_r$ denote the intrinsic spins of $D$ and $r$, and $L$ is the relative orbital angular momentum between $r$ and $c$.

Relativistic {\it Breit-Wigner} functions are phenomenological descriptions of non-overlapping intermediate transitions that are away from threshold. Their dynamical structure takes the form
\begin{equation}
\label{eq:bw}
T_r(s) = \frac{1}{m_0^2 - s - im_0\Gamma(s)},
\end{equation}
where $m_0$ is the resonance mass and $\sqrt{s}$ is the resonance two-particle invariant mass. The momentum-dependent resonance width $\Gamma(s)$ relates to the pole width ($\Gamma_0$) as
\begin{equation}
\label{eq:gamma_s}
\Gamma(s) = \Gamma_0 \frac{m_0}{\sqrt{s}} \left( \frac{q}{q_0} \right)^{(2L+1)} \mathcal{B}_r^{L}(q,q_0).
\end{equation}
Pole masses and widths in this analysis are fixed to the PDG values~\cite{ref:pdg20}. The function $\mathcal{B}_r^L(q,q_0)$ is the centrifugal-barrier factor~\cite{ref:barrierhippel} in the decay $r \to ab$, where $q$ is the momentum transfer in the $r$ decay in its rest frame and $q_0$ is $q$ evaluated at $m_0$. The full Breit-Wigner amplitude description includes, in addition to the dynamical part $T_r(s)$, barrier factors corresponding to $P$- and $D$-wave decays of the initial state $D$ meson and resonance $r$ decays, $\mathcal{B}_{D}^L$ and  $\mathcal{B}_r^L$ respectively, and an explicit spin-dependent factor ($\mathcal{Z}_L$):
\begin{equation}
\label{eq:amp_bw}
A_r^{BW} = T_r(s) \times \mathcal{B}_{r}^L(q,q_0) \times \mathcal{B}_D^L(p,p_0) \times \mathcal{Z}_L(J_D,j_r,{\bf p},{\bf q}). 
\end{equation}
Here $p$ denotes momentum of the spectator particle $c$ in the resonance rest frame and $p_0$ is the corresponding on-shell value.
Scaling of the Breit-Wigner lineshape by the barrier factors optimizes the enhancement or dampening of the total amplitude depending upon the relative orbital angular momentum (or the spin of the resonance) of the decay and the linear momenta of the particles involved. For resonances with spin greater than or equal to one and small decay interaction radius (or impact parameter) of the order 1~fm, large momentum transfer in the $a,b$ system is disfavoured because of limited orbital angular momentum between $r$ and $c$. {\it Blatt-Weisskopf} form factors~\cite{ref:blattweisskopf}, normalized to unity at $q=q_0$, are used to parametrize the barrier factors whose functional forms are given in table~\ref{tab:blatt}, where $d$ denotes the interaction radius of the parent particle. Similar expressions for $D$-decay barrier factors can be written in terms of momentum of the spectator particle evaluated in the $D$ rest frame.

\begin{table}[tbp]
\centering
\large
\begin{tabular}{|c|c|}
\hline
         $L$ & Form factor $\mathcal{B}^L_r(q,q_0)$ \Tstrutsmall \Bstrutsmall  \\
\hline \hline
         0   & 1  \\
         
         1   & $\sqrt{\frac{1 + q_0^2d^2}{1 + q^2d^2}}$ \Tstrut \Bstrut \\
         
         2   & $\sqrt{\frac{9 + 3q_0^2d^2 + (q_0^2d^2)^2}{9 + 3q^2d^2 + (q^2d^2)^2}}$ \Tstrut \Bstrut \\ 
\hline
\end{tabular}
    \caption{Normalized Blatt-Weisskopf barrier factors~\cite{ref:blattweisskopf} for resonance decay exhibiting spin and momentum-dependent effects.}
    \label{tab:blatt}
\end{table}
The spin-dependence of the decay amplitudes are derived using covariant spin-tensor or Rarita-Schwinger formalism~\cite{ref:chungcernreport, ref:spinchung, ref:spintensor, ref:spinzoubugg}. The pure spin-tensors for spin 1 and 2 from spin-projection operators $\Theta$ and the break-up four-momentum $k^{\mu} = a^{\mu} - b^{\mu}$ in the resonance rest frame (so that three-momentum \textbf{k} = 2\textbf{q}) are given by

\begin{align}
    \mathit{S}_{\mu} &= \Theta_{\mu \nu}k^{\nu}, \\
    \mathit{T}_{\mu \nu} &= \Theta_{\mu \nu \rho \sigma}k^{\rho}k^{\sigma}.
\end{align}
Using the above defined spin-tensors together with the orthogonality and spacelike conditions on Lorentz invariant functions of rank one and two for $P$ and $D$ wave, respectively, when summed over all the polarization states, it is possible to arrive at the following definitions of the angular decay amplitude:

\begin{align}
    A(0 \to 1 + 1) &: \mathcal{Z}_{L=1} = \Theta^{\mu \nu} \mathit{S}_{\mu} \mathcal{L}_{\nu}, \\
    A(0 \to 2 + 2) &: \mathcal{Z}_{L=2} = \Theta^{\mu \rho}\Theta^{\nu \sigma} \mathit{T}_{\mu \nu} \mathcal{M}_{\rho \sigma},
\end{align}
where $\mathcal{L}$ and $\mathcal{M}$ are normalized tensors describing the states of relative orbital angular momenta $L=1$ and $L=2$. Simplified expressions for the angular amplitudes in terms of four-momenta of the states involved and their invariant masses are given in appendix~\ref{appensec:BW_spin}. 

Overlapping $S-$wave pole production with multiple channels in two-body scattering processes are best described by the $K$-matrix formulation~\cite{ref:KmatChung, ref:kmatparam, ref:focuskmat}. A sum of Breit-Wigner functions to describe such broad resonant structures violate the unitarity of the transition matrix $\mathcal{T}$. The idea is to write the total effective $\mathcal{T}$ matrix in terms of the $K$ matrix
\begin{equation}
    \hat{\mathcal{T}} = ( 1 - i\hat{K}\omega )^{-1}\hat{K}.
\end{equation}
The $K$-matrix contains contributions from all the poles, intermediate channels and all possible couplings. Here the parameter $\omega$ is a diagonal matrix with the phase-space densities of various channels involved as its elements. As an example, in low energy $\pi\pi \to \pi\pi$ scattering, the total amplitude carries a contribution from coupling between the resonance $f_0(980)$ and a $KK$ channel, which is partly responsible for the sharp dip observed in the scattering amplitude near $1$ GeV. A simple Breit-Wigner function cannot explain this variation in the amplitude. 

This recipe can be translated to decay processes involving broad overlapping resonance structures produced in an $S-$wave. An initial state first couples to $K$-matrix poles with strength parametrized by $\beta_{\alpha}$ for pole $\alpha$, and these poles in turn couple to various intermediate channels $i$ in the $K$-matrix with strengths characterized by $g_{i}^{\alpha}$. Direct coupling between initial state and these $K$-matrix channels is also a possibility and the corresponding strength is denoted by $f_{1i}^{\rm prod}$, scaling a slowly varying polynomial term in $s$ and an arbitrary parameter $s_0^{\rm prod}$, which are fixed from a global analysis of $\pi\pi$ scattering data~\cite{ref:kmatparam}. Summing these contributions together results in the {\it production vector} $\hat{P}$: 
\begin{equation}
\hat{P}_i = \sum_{\alpha} \frac{\beta_{\alpha}g_i^{\alpha}}{m_{\alpha}^2-s} + f_{1i}^{\rm prod}\frac{1-s_0^{\rm prod}}{s-s_0^{\rm prod}},
\end{equation}
where $m_{\alpha}$ are the pole masses and $s$ is the kinematic variable, in this case the invariant squared-mass of the two pions from the three-body decay. The structure of the $K$ matrix in a decay process with poles $\alpha$ and decay channels denoted by $i$ and $j$ is given by
\begin{equation}
\label{Eq:kmat_def}
K_{ij}(s) = \left( \sum_{\alpha} \frac{g_i^{\alpha}g_j^{\alpha}}{m_{\alpha}^2-s} + f_{ij}^{\rm scatt}\frac{1-s_0^{\rm scatt}}{s-s_0^{\rm scatt}} \right) f_{A0}(s).
\end{equation}
The intermediate channels considered in the present case are $\pi\pi$, $KK$, $\pi\pi\pi\pi$, $\eta\eta$ and $\eta\eta'$. In addition to the pole terms, direct scatterings between channels are also considered with strengths $f_{ij}^{\rm scatt}$ in a polynomial term in $s$ and parameter $s_0^{\rm scatt}$~\cite{ref:kmatparam}.
An arbitrary kinematic singularity appears below the $\pi\pi$ production threshold at $\sqrt{s}~\sim ~ m_{\pi}/\sqrt{2}$. The so-called {\it Adler-zero} term, $f_{A0}(s)$~\cite{ref:bellebabar18} is multiplied to the entire $K$-matrix to suppress it. Finally, the total production amplitude for a final decay channel $j$ can be written in terms of the $P$ vector as
\begin{equation}
\label{Eq:prod_amp}
A_{\pi\pi}^{(L=0)}(s)_j = \left( I - i\hat{K}(s)\hat{\omega}(s) \right)^{-1}_{ji} P_{i}(s).
\end{equation}

\begin{table}[tbp]
\centering
\begin{tabular}{|c|c|}
\hline
$\alpha$ & $m_{\alpha}$ [GeV/$c^2$] \Tstrut \Bstrut \\
\hline \hline
1	&	0.65100 \\
2	&	1.20360	\\
3	&	1.55817 \\
4	&	1.21000 \\
5	&	1.82206 \\
\hline
\end{tabular}
\caption{Poles and couplings in $K$-matrix \cite{ref:bellebabar18, ref:kmatparam}. The intermediate channels include $\pi\pi$, $KK$, $\pi\pi\pi\pi$, $\eta\eta$ and $\eta\eta'$}
\label{tab:kmat_poles}
\end{table}
In the present case of $D^0 \to K^0(\pi\pi)_{S}$, five poles of $K$ matrix are considered, which are summarized in table~\ref{tab:kmat_poles}. These may be associated with $\mathcal{T}$ poles as physical resonances: $f_0(980),~f_0(1370),~f_0(1500),~f_0(1710),$ and a broad spectrum $f_0(1200-1600)$. Moreover, $K$ matrix couplings associated with only $\pi\pi$ final states, {\it i.e.}, $j=1$ or first row of the $( I - i\hat{K}\hat{\omega} )^{-1}$ matrix, are considered.  

The $K\pi$ scalar contribution is described by a parametrization developed by the LASS collaboration, again originally designed for scattering processes~\cite{ref:lassdef}. The first scalar excitation of the $K\pi$ state is $K_{0}^{*}(1430)$, so the CF non-resonant process carries a considerably larger contribution and is described by the empirical LASS formulation. The total LASS amplitude is a sum of a Breit-Wigner resonant term and a non-resonant scattering term, scaled by an overall complex coupling parameter as
\begin{equation}
A_{K\pi}^{(L=0)}(s) = a_re^{i\phi_r}(R\sin\delta_Re^{i\delta_R}e^{i2\delta_S} + S\sin\delta_Se^{i\delta_S}),
\end{equation}
where
\begin{align*}
    \delta_R &= \phi_R + \tan^{-1} \left[ \frac{m_0\Gamma(s)}{m_0^2 - s} \right], \\
    \delta_S &= \phi_S + \cot^{-1}\left[ \frac{1}{aq} + \frac{r_{\rm int}q}{2} \right] .
\end{align*}
\begin{figure}[b!]
\centering
\includegraphics[scale=0.45]{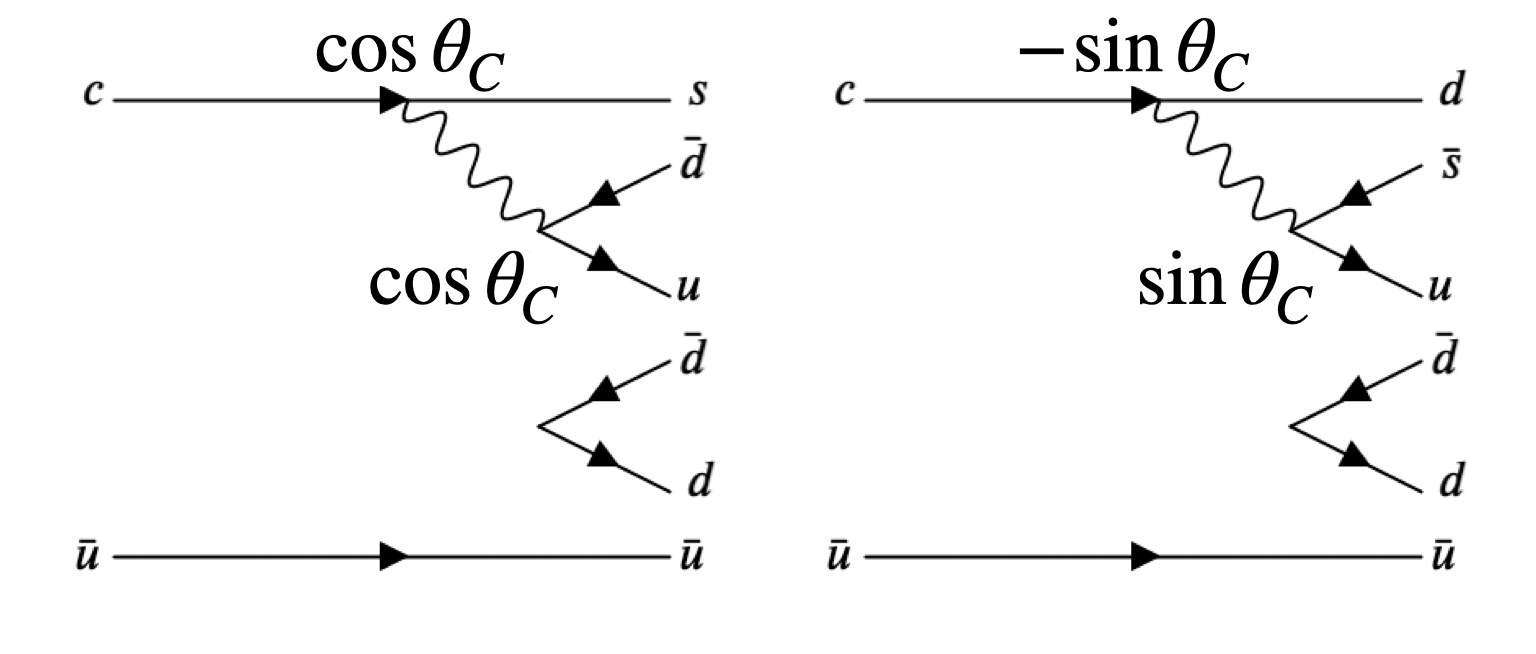}
\caption{\label{fig:cf_dcs_fd} CF $D^0 \to \bar{K}^0\pi^+\pi^-$ (\textit{left}) and DCS $D^0 \to K^0\pi^+\pi^-$ (\textit{right}) decay diagrams.}
\end{figure}
The parameters $Re^{i\phi_R}$ and $Se^{i\phi_S}$ are relative complex amplitudes of the resonant and non-resonant (direct scattering) terms, respectively. 
The parameter $a$ is the scattering length and $r_{\rm int}$ is the effective interaction length in the case of direct scattering. More details can be found in Ref.~\cite{ref:lassdef}. The same parametrization is used for the DCS $K_0^{*}(1430)^+$ resonant process for which $a_r$ and $\phi_r$ are again nominally to be determined from the fit.

In the remainder of this section we discuss the application of the isobar model to $D^0\to K^0_{\rm S,L}\pi^+\pi^-$ decays. Non-trivial effects on the rate of hadronic decays involving pions and neutral kaons as a result of DCS transitions interfering with CF transitions are expected~\cite{ref:bigiyamamoto}. A manifestation of this interference effect is a small difference in the decay rates of processes involving a $K_{\rm S}^0$ versus a $K_{\rm L}^0$ in the final state. Consider the decay process $D^0 \to K_{\rm S}^0\pi^+\pi^-$, which can proceed both via the CF $D^0 \to \bar{K}^0\pi^+\pi^-$ and the DCS $D^0 \to K^0\pi^+\pi^-$.

In addition to the pure CF and DCS sub-transitions, as shown in figure~\ref{fig:cf_dcs_fd}, a mixture of these, where the two pions are produced in a $CP$ eigenstate resonance, is also a viable transition process to the $K_{\rm S}^0\pi^+\pi^-$ final state. Note that the two amplitudes would be identical under the interchange of the $s$ and $d$ quarks involved in the weak interaction, which is referred to as U-spin symmetry. Using the phase convention $CP\ket{K^0} = - \ket{\bar{K}^0}$, the partial amplitude of intermediate processes involving only neutral $CP$ eigenstate resonances $k_{CP}$, \textit{i.e.} $D^0 \to K_{\rm S}^0(\pi^+\pi^-)_{k_{CP}}$, can be written as a superposition of $D^0 \to \bar{K}^0\pi^+\pi^-$ and $D^0 \to K^0\pi^+\pi^-$ as follows

\begin{subequations}
\begin{align}
A \left( D^0 \to K_{\rm S}^0(\pi\pi)_{k_{CP}} \right) &= \frac{1}{\sqrt{2}} \left( A(D^0 \to \bar{K}^0(\pi^+\pi^-)) - A(D^0 \to K^0(\pi^+\pi^-)) \right) \\
								    &= \frac{1}{\sqrt{2}} A(D^0 \to \bar{K}^0(\pi^+\pi^-)) \left( 1 - \frac{A(D^0 \to K^0(\pi^+\pi^-))}{A(D^0 \to \bar{K}^0(\pi^+\pi^-))} \right). 
\end{align}
Defining 
$  A(D^0 \to K^0(\pi^+\pi^-))/A(D^0 \to \bar{K}^0(\pi^+\pi^-)) = -{\rm tan}^2\theta_{\rm C}~\hat{\rho},$ where $\hat{\rho}$ may be understood as a \textit{U-spin breaking} parameter, the $K_{\rm S}^0\pi^+\pi^-$ $CP$ transition amplitude can be written as  
\begin{equation}
A(D^0 \to K_{\rm S}^0(\pi\pi)_{k_{CP}}) = \frac{1}{\sqrt{2}}A(D^0 \to \bar{K}^0(\pi^+\pi^-)) \left( 1 + ~\hat{\rho}_{k_{CP}}~{\rm tan}^2\theta_C \right).
\end{equation}
\end{subequations}
The U-spin breaking parameter, $\hat{\rho}$ (= $re^{i\delta}$) is a complex and purely empirical quantity. For a three-body decay with multiple exclusive $CP$ eigenstate resonant contributions ($k_{CP}$), $\hat{\rho}$ factors for each are denoted by $\hat{\rho}_{k_{CP}}$. The $\hat{\rho}$ parameters carry the phase-shifts generated as a result of DCS interference and naively, the magnitudes $|\rho|$ are expected to be $\sim \mathcal{O}(1)$ in the absence of any interference between CF and DCS transitions. However, the magnitudes should be empirically measured to consider the possibility of deviation from the nominal ${\rm tan}^2\theta_{\rm C}$ Cabibbo factor. 
A similar treatment for the decay process $D^0 \to K_{\rm L}^0\pi^+\pi^-$ yields

\begin{subequations}\label{eq:y}
\begin{align}
	A(D^0 \to K_{\rm L}^0(\pi\pi)_{k_{CP}}) &= \frac{1}{\sqrt{2}}A(D^0 \to \bar{K}^0(\pi^+\pi^-)) \left( 1 - ~\hat{\rho}_{k_{CP}}~{\rm tan}^2\theta_{\rm C} \right).
\end{align}
\end{subequations}
It is then straightforward to show that the $CP-$resonant amplitudes of $D^0 \to K_{\rm L}^0\pi^+\pi^-$ decay process can be related to the $D^0 \to K_{\rm S}^0(\pi^+\pi^-)_{k_{CP}}$ amplitudes as
\begin{equation}
\frac{A(D^0 \to K_{\rm L}^0(\pi\pi)_{k_{CP}})}{A(D^0 \to K_{\rm S}^0(\pi\pi)_{k_{CP}})} = \frac{1-{\rm tan}^2\theta_{\rm C}~\hat{\rho}_{k_{CP}}}{1+{\rm tan}^2\theta_{\rm C}~\hat{\rho}_{k_{CP}}} \approx 1 - 2{\rm tan}^2\theta_{\rm C}~\hat{\rho}_{k_{CP}} + \mathcal{O}({\rm tan}^4\theta_{\rm C}) , \nonumber
\end{equation} 
which results in the relation 
\begin{equation}
\label{eq:master}
A(D^0 \to K_{\rm L}^0(\pi\pi)_{k_{CP}}) = \left( 1 - 2{\rm tan}^2\theta_{\rm C}~\hat{\rho}_{k_{CP}} \right) \times A(D^0 \to K_{\rm S}^0(\pi\pi)_{k_{CP}}),
\end{equation}
where terms higher than second order in ${\rm tan}\theta_{\rm C}$ are neglected. An amplitude model description of the $D^0 \to K_{\rm L}^0\pi^+\pi^-$ decay mode is required for a constrained strong-phase measurement. The $K_{\rm L}^0\pi^+\pi^-$ amplitude model can be obtained via the DCS interference motivated modifications to the well studied $K_{\rm S}^0\pi^+\pi^-$ model, such as the one stated in eq.~\ref{eq:master}. Another departure is expected in the DCS resonant modes such as $D^0 \to (K_{\rm L,S}^{0}\pi^+)_{K^{*}}\pi^-$, with a relative minus sign between $K_{\rm S}^0\pi^+\pi^-$ and $K_{\rm L}^0\pi^+\pi^-$ amplitudes, again because of the phase structure in the definition of $K_{\rm L,S}^{0}\pi^+\pi^-$ in terms of the flavor states. We insert this minus sign in the DCS amplitudes of $K_{\rm L}^0\pi^+\pi^-$ instead of $K_{\rm S}^0\pi^+\pi^-$ to maintain consistency with the standalone $K_{\rm S}^0\pi^+\pi^-$ amplitude model that has no minus sign in the DCS parts. Doing so merely introduces an extra $180^{\circ}$ phase added to the nominal DCS phases and does not affect any physics. The total amplitudes are

\begin{align}
A(D^0 \to K_{\rm S}^0\pi^+\pi^-) &= \sum_{r}A^{CF}_{\bar{K}^0\pi\pi} + \sum_{r\prime}A^{DCS}_{K^0\pi\pi} + \sum_{k_{CP}}A^{CP}_{K_{\rm S}^0(\pi^+\pi^-)_{k_{CP}}} , \label{Eq:kspipi_amp}\\
A(D^0 \to K_{\rm L}^0\pi^+\pi^-) &= \sum_{r}A^{CF}_{\bar{K}^0\pi\pi} - \sum_{r\prime}A^{DCS}_{K^0\pi\pi} + \sum_{k_{CP}} \left( 1 - 2{\rm tan}^2\theta_C~\hat{\rho}_{k_{CP}} \right) A^{CP}_{K_{\rm S}^0(\pi^+\pi^-)_{k_{CP}}}. \label{Eq:klpipi_amp}
\end{align}
The only way to determine the $\hat{\rho}$ parameters associated with each of the two-body intermediate resonant structure contributions in $K_{\rm S,L}^0\pi^+\pi^-$ decay process, is to fit an amplitude model for $D^0 \to K_{\rm L}^0\pi^+\pi^-$, where the $D^0 \to K_{\rm S}^0\pi^+\pi^-$ decay may be used as a constraint in a simultaneous fit.

\section{BESIII detector and simulated sample}
\label{sec:detdata}

The BESIII detector~\cite{ref:bes3exp} records symmetric $e^+e^-$ collisions provided by the BEPCII storage ring~\cite{ref:bepc2}, which operates with a peak luminosity of $1\times10^{33}$~cm$^{-2}$s$^{-1}$ in the center-of-mass energy range from 2.0 to 4.99~GeV. BESIII has collected large data samples in this energy region~\cite{ref:Ablikim2019hff}. The cylindrical core of the BESIII detector covers 93\% of the full solid angle and consists of a helium-based multilayer drift chamber~(MDC), a plastic scintillator time-of-flight system~(TOF), and a CsI(Tl) electromagnetic calorimeter~(EMC), which are all enclosed in a superconducting solenoidal magnet providing a 1.0~T magnetic field. The solenoid is supported by an octagonal flux-return yoke with resistive plate counter muon identification modules interleaved with steel. The charged-particle momentum resolution at $1~{\rm GeV}/c$ is $0.5\%$, and the ${\rm d}E/{\rm d}x$ resolution is $6\%$ for electrons from Bhabha scattering. The EMC measures photon energies with a resolution of $2.5\%$ ($5\%$) at $1$~GeV in the barrel (end cap) region. The time resolution in the TOF barrel region is 68~ps, while that in the end cap region is 110~ps.

The experimental data used were collected with a centre-of-mass energy corresponding the mass of the $\psi(3770)$ resonance. The sample size corresponds to an integrated luminosity of 2.93~fb$^{-1}$. 

We also use simulated events to optimize our selection, identify background contributions and validate the amplitude analysis. In the BESIII software framework, starting from $e^+e^-$ annihilation upto the charmonium resonance production part of the processes, including the initial-state radiation (ISR) effects and the beam energy spread of 0.97 MeV, are simulated using the \textsc{kkmc} generator~\cite{ref:kkmc} and for the resonance decay, elaborate \textsc{BesEvtGen} models~\cite{ref:besevtgen} are used for they also contain dynamical information of the decay. The resonances supported by \textsc{kkmc} include $J/\psi$, $\psi(2S)$, $\psi(3770)$, $\psi(4040)$, $\psi(4160)$, $\psi(4415)$ and other low lying resonances like $\rho$, $\phi$, $\omega$ and their excitations. 

Both inclusive and signal samples of simulated events are produced using the above mentioned generator packages, as well as a \textsc{Geant4}~\cite{ref:geant4} - based detector geometry and response simulation package. The inclusive simulation sample in this analysis is prepared by adding together various simulated physics processes in proportion to their branching ratios. These physics processes include $D^0\bar{D}^0$ and $D^+D^-$ from $\psi(3770)$, $J/\psi$ and $\psi(2S)$ charmonium production along with ISR, lepton pair production and $q\bar{q}$ continuum. The size of the inclusive simulation sample used for background estimation is roughly ten times that of the experimental data. Simulated samples of \kspipi\ and \klpipi\ decays, with a size one hundred times that of experimental data, are produced to normalize the probability density in the amplitude fit. Simulated signal decays including resonant structures are produced to validate the amplitude fit.

\section{Event selection}
\label{sec:evtsel}
We use a sample of quantum-correlated $e^+e^-\to D^0\bar{D}^0$ events, which are produced close to the kinematic threshold for this process. No other particles accompany the $D$ mesons, which results in a low-background environment to reconstruct the $D$ candidates. We identify the flavor of the neutral $D$ meson decaying into the signal modes $K_{\rm S,L}^0\pi^+\pi^-$ by reconstructing the other $D$ meson state in a flavor-specific mode, which is also commonly known as the {\it tag mode}, and this full-reconstruction technique is referred to as the {\it double-tag method}. The \klpipi\ signal mode is reconstructed with the \kl\ candidate treated as a {\it missing particle}, which makes using the semi-leptonic exact flavor-tag modes with high branching fraction such as $K^+e^-\bar{\nu}_e$ infeasible. Therefore, $K^+\pi^-$, $K^+\pi^-\pi^+\pi^-$ and $K^+\pi^-\pi^0$ hadronic flavor tag modes are utilized in this analysis to select the signal decay modes \dtokslpipi. We account for the small DCS contamination of these hadronic flavor tags as part of the analysis. Note that inclusion of charge-conjugate processes is implied throughout unless stated otherwise. 

Charged particles are reconstructed in the tracking system within the MDC acceptance $|{\rm cos}\theta|<0.93$, where $\theta$ is the polar angle of the track with respect to the axis of the MDC ($z$-axis). For the charged particles that are direct products of the the $D$ mesons, we require the distance of closest approach to the interaction point (IP) to be less that 1~cm in the $x-y$ plane and less than 10~cm along the $z$-axis. Whereas for the pion candidates used to reconstruct $K^0_{\rm S}\to\pi^+\pi^-$ decays, the only condition is on their distance to the IP along the $z$-axis, which is required to be less than 20~cm.  We identify charged particles (PID) using combined probabilities from both time-of-flight information from the TOF and ${\rm d}E/{\rm d}x$ measurements from the MDC under the pion and kaon hypotheses. The hypothesis with the greater combined probability is chosen and the charged particle is identified as a pion or a kaon accordingly. 

To select photon candidates from showers in the EMC, we require energy deposits of at least 25~MeV in the barrel region of the EMC ($|\!\cos\theta| < 0.8$)  and at least 50~MeV in the end-cap region ($0.86 < |\!\cos\theta| < 0.92$). Photon candidates must also be isolated from every charged track in an event by more than $10^{\circ}$ to suppress the hadron interaction induced clusters in the EMC. Furthermore, to suppress clusters associated with either beam background or electronic noise, we require the time elapsed between the bunch crossing and the cluster's detection in the EMC is less than 700~ns. To reconstruct $\pi^0$ candidates, it is required that the invariant mass of a pair of photons lies within (0.110, 0.155) GeV$/c^2$. For better resolution, a kinematic fit is performed to constraint the di-photon mass to the nominal $\pi^0$ mass and the corresponding output four-momentum is utilized in the analysis.

Further selection is performed to suppress combinatorial backgrounds. We use two kinematic variables to identify tag and signal $D$ mesons: the \textit{energy difference} $\Delta E = \sqrt{s}/2 - E_D$ and the {\it beam-energy-constrained mass},
\begin{equation}
    M_{\rm BC} = \sqrt{(\sqrt{s}/2)^{2} - \sum_i | \mathbf{p}_i|^2 }, \label{eq:mbc}
\end{equation}
where $E_D$ is the measured $D$ meson energy and $\mathbf{p}_i$ denotes the momentum vector of the $i^{\rm th}$ final state particle of the $D$ meson under study. Signal decays peak at zero and the known $D$ mass in the $\Delta E$ and $M_{\rm BC}$ distributions, respectively, whereas combinatorial background does not peak at all. The signal peak in the $\Delta E$ and $M_{\rm BC}$ distributions for the three tag-modes, $K^+\pi^-$, $K^+\pi^-\pi^+\pi^-$ and $K^+\pi^-\pi^0$ are modeled with double-Gaussian functions. Background distributions are described with polynomial and Argus functions~\cite{ref:argus} in the $\Delta E$ and $M_{\rm BC}$ distributions, respectively. Candidate $D$ mesons are required to fall within intervals that are $\pm 3\sigma$ about the signal peaks of both $\Delta E$ and $M_{\rm BC}$ distributions. For events containing multiple tag-side $D$ candidates satisfying all the conditions mentioned thus far, the combination with minimum $|\Delta E|$ is selected. To suppress the cosmic ray, Bhabha and di-muon background events in the $K^+\pi^-$ tag mode, two charged tracks, neither identified as an electron nor muon, with TOF time difference less than 5~ns are required. The $K^+\pi^-\pi^+\pi^-$ tag-mode decays contain a peaking background from $K_{\rm S}^0K^+\pi^-$ candidates with about $2\%$ contamination rate, as estimated from the inclusive simulation sample. To suppress this background, a $K_{\rm S}^0$ mass veto within the range [0.479, 0.518] GeV$/c^2$ is applied on both permutations of oppositely charged pions selected in the $K^+\pi^-\pi^+\pi^-$ final state, reducing the $K_{\rm S}^0K^+\pi^-$ background to a negligible level of $0.09\%$. 

For the \kspipi\ signal mode, the number of charged particles passing all the conditions, apart from those used to reconstruct the tag decay, is required to be greater than or equal to four in an event. Candidate \ks\ selection is performed in three steps while examining all possible combinations of the four selected charged tracks. Firstly, successive vertex fits are performed on the primary pions from $D$ meson and the pair of pions being tested as final state particles coming from \ks decays. The second step entails enforcing a \ks\ mass window condition with $3\sigma$ bounds [0.485, 0.510]~GeV$/c^2$ on the the invariant mass of a pair of oppositely charged pions. Finally, a flight-significance criterion is placed wherein the decay length of the \ks\ candidate is required to be greater than twice its uncertainty. Pion tracks that are used to reconstruct \ks\ candidates are not required to satisfy the particle identification criteria. Furthermore, events with multiple \ks\ candidates are rejected to remove $D\to$\ks\ks$X$ decays. 

To improve the momentum resolution, kinematic fits with all the final state particle momenta from both tag and signal sides are performed and the events for which the fit does not converge are discarded. Constraints related to the total four-momentum, $D$ and \ks\ masses are put in place. The signal efficiencies of the kinematic fit selection criteria for $K^+\pi^-$, $K^+\pi^-\pi^+\pi^-$ and $K^+\pi^-\pi^0$ tagged \kspipi\ mode are $97.2\%$, $95.2\%$ and $93.7\%$ respectively.

\begin{figure}[tbp]
    \centering
    \includegraphics[scale=0.36]{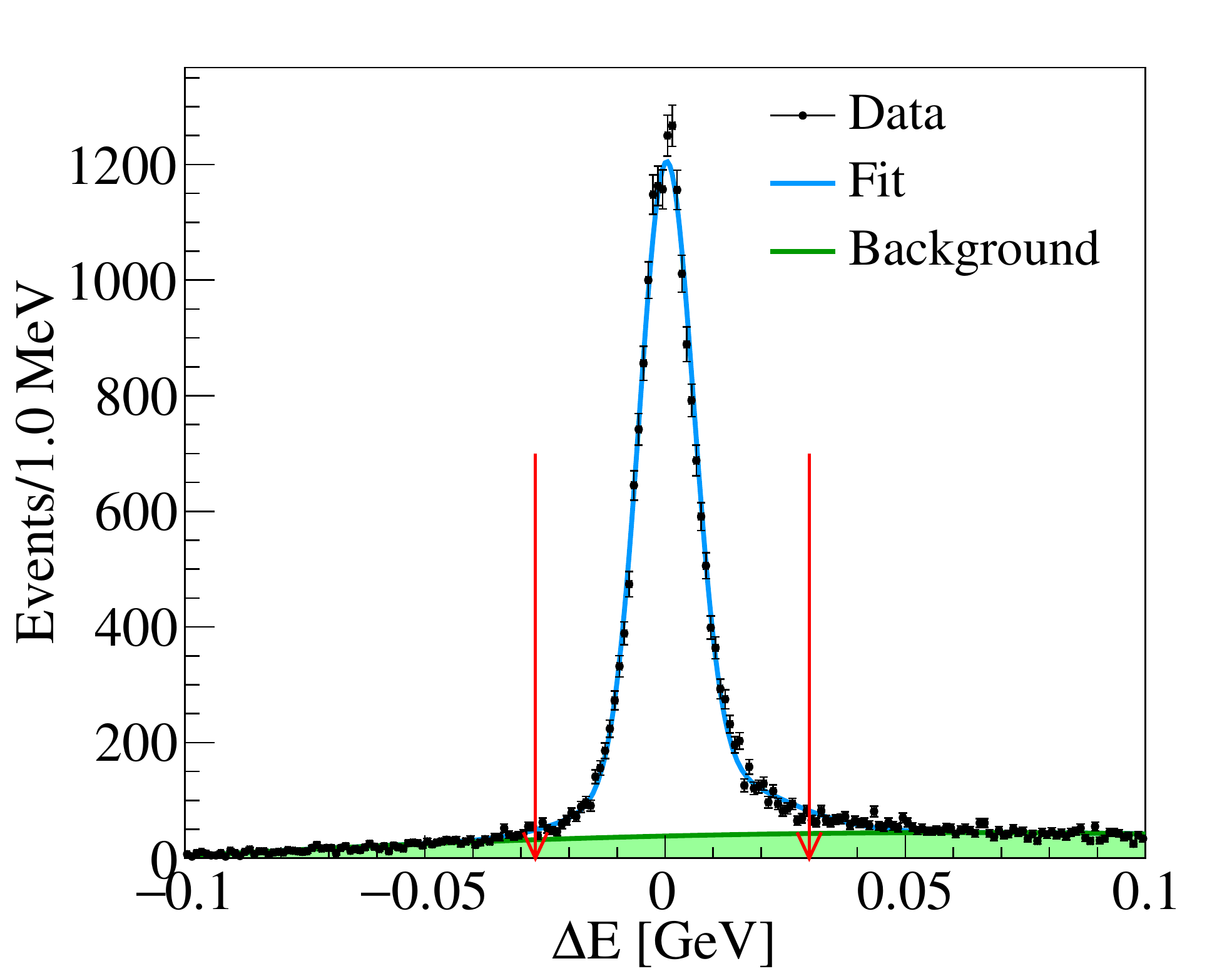}
    \includegraphics[scale=0.36]{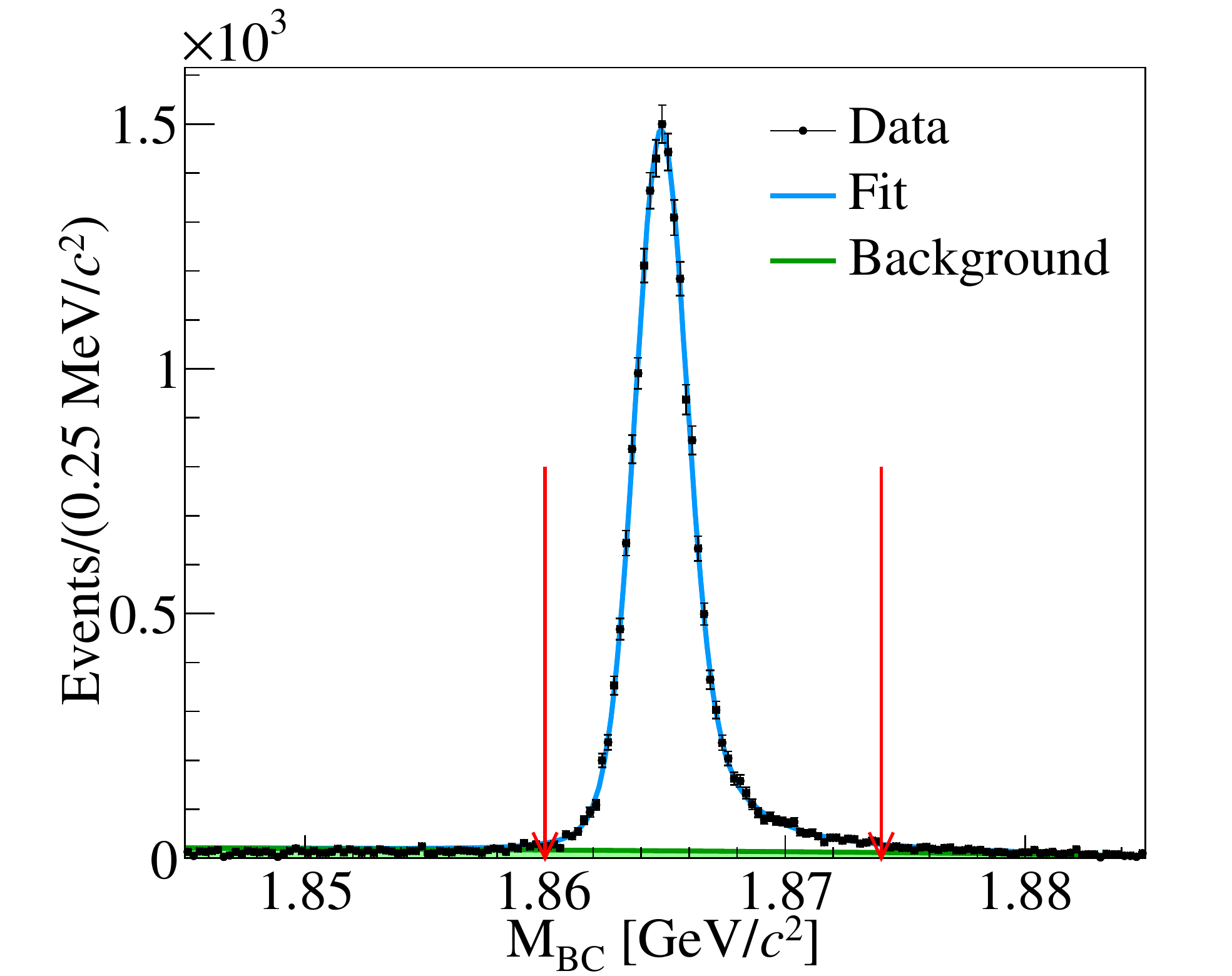}\\
    \includegraphics[scale=0.36]{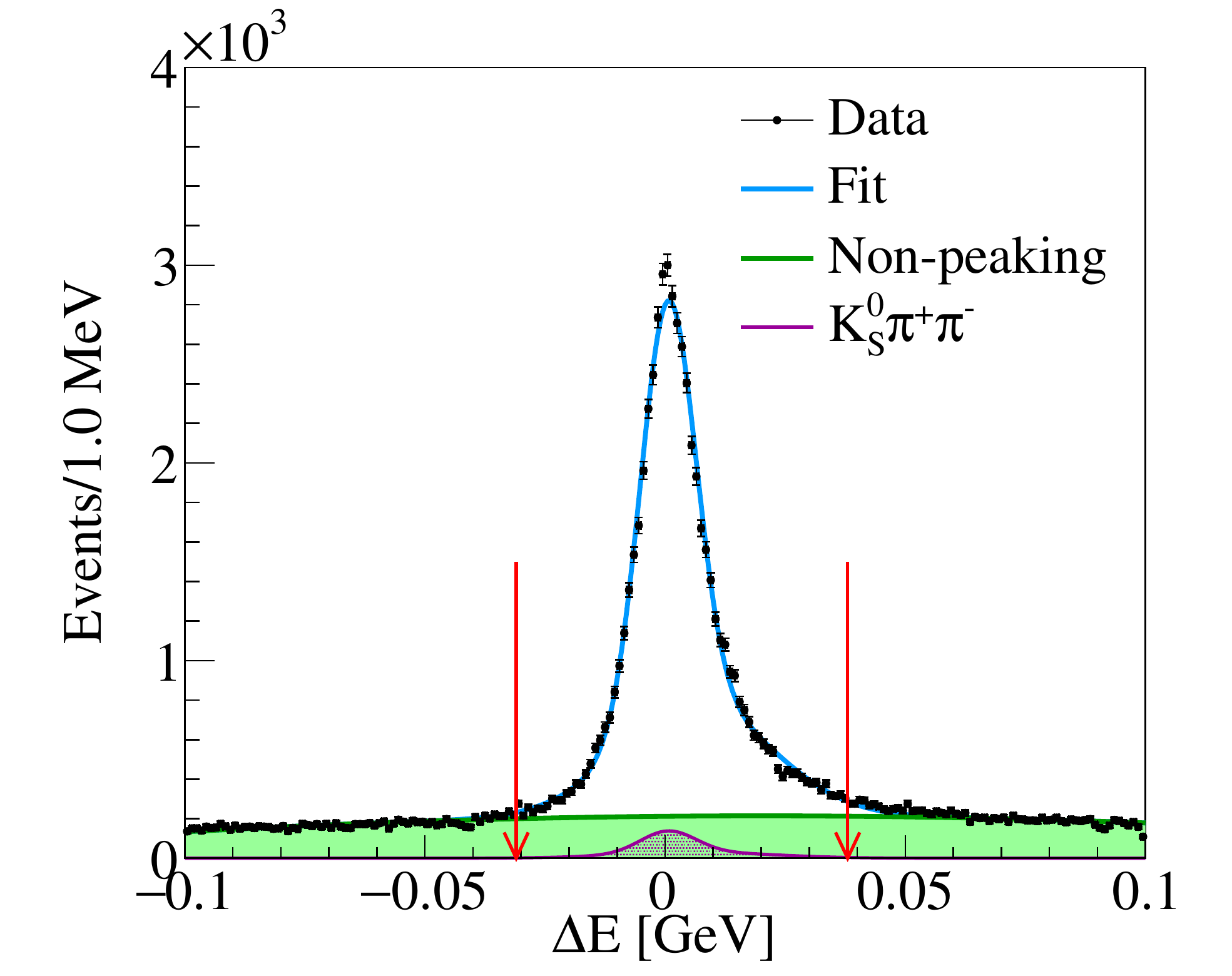}
    \includegraphics[scale=0.36]{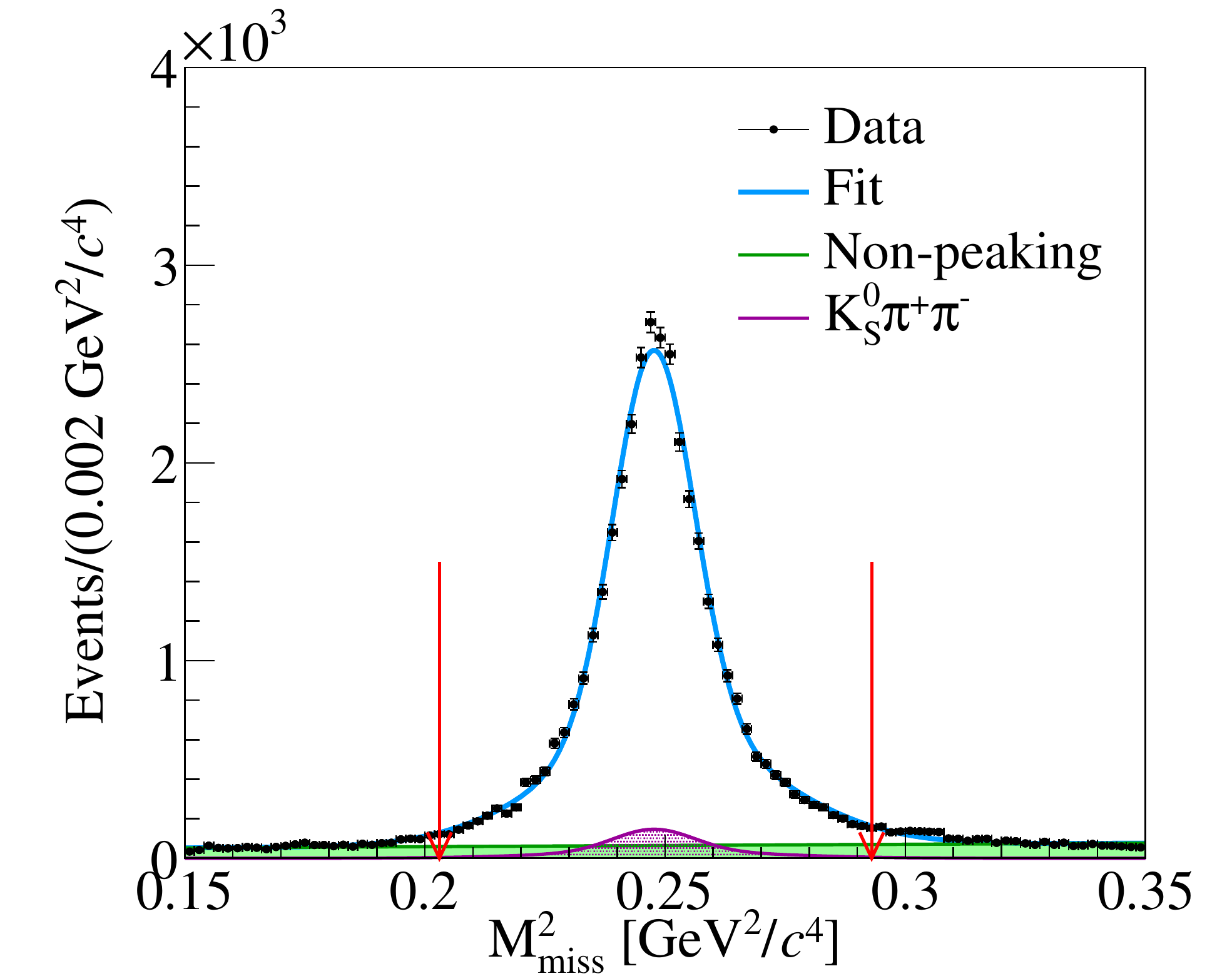}
    \caption{The $\Delta E$, $M_{\rm BC}$ distributions for \kspipi\ signal candidates ({\it top}) and $\Delta E$, M$_{\rm miss}^2$ distributions for \klpipi\ signal candidates ({\it bottom}) in data. The red arrows indicate the allowed signal regions.}
    \label{fig:sig_deltae_mbc}
\end{figure}

In the \klpipi\ signal selection, to suppress the \kspipi\ peaking background, we require the number of charged particle tracks, that are not used to reconstruct the tag, to be exactly two, both of which must satisfy the above mentioned conditions of track selection. The residual four-momentum in the detector, called {\it missing-momentum}, after reconstructing all the charged tracks on the tag side and both the pion tracks on the signal side, is identified as a \kl\ candidate; this method is referred to as the \textit{missing-mass technique}. 
For both the signal modes, events containing a $\pi^0$ or $\eta$ candidate are vetoed for which the invariant mass of any permutation of pairs of photons falls in the respective mass range of [$0.095, 0.165$] GeV$/c^2$ and [$0.48, 0.58$] GeV$/c^2$. For the $K_{\rm L}^0\pi^+\pi^-$ mode, the $\pi^0$ veto removes a significant fraction of the $D^0\to K^0_{\rm S}\pi^+\pi^-$ background, where $K_{\rm S}^0\to \pi^0\pi^0$. To determine the selection criteria on $M_{\rm BC}$ and $\Delta E$ for the $D\to K^0_{\rm S}\pi^+\pi^-$ decays, we model both the distributions by performing
 maximum likelihood fits as shown in figure~\ref{fig:sig_deltae_mbc}. The signal part in both $\Delta E$ and $M_{\rm BC}$ distributions is described by double Gaussian functions. The combinatorial background in the $\Delta E$ and $M_{\rm BC}$ distributions is modeled with polynomial and Argus functions, respectively. For the \klpipi\ signal mode, a distribution of missing-mass squared defined as
\begin{equation}
    {\rm M_{miss}^2} = (\sqrt{s}/2 -  E_{\pi^+}-E_{\pi^-})^2 - |\mathbf{p}_{\rm tag}+\mathbf{p}_{\pi^+}+\mathbf{p}_{\pi^-}|^2 ,
\end{equation}
is analysed, where ($E_{\pi^{\pm}}$, $\mathbf{p}_{\pi^{\pm}}$) is the four-momentum of $\pi^{\pm}$ candidates on the signal side and $\mathbf{p}_{\rm tag}$ is the total momentum of the single-tag $D$ meson. The M$_{\rm miss}^2$ distribution is modeled with double Gaussian functions for signal and peaking background, and a straight line for the combinatorial background. Candidates beyond a $3\sigma$ coverage about the mean, the bounds of which are given in table~\ref{tab:sig_var}, are rejected in all the three kinematic variables. The total yields obtained after full reconstruction and selection are 16490 for the \kspipi\ mode and 39085 for the \klpipi\ mode. 

\begin{table}[tbp]
\centering
\begin{tabular}{|c|c|c|}
\hline
Variable & $K_{\rm S}^0\pi^+\pi^-$ & $K_{\rm L}^0\pi^+\pi^-$ \Tstrut \Bstrut \\
\hline \hline
$\Delta E$ [GeV]       & $[-0.027,0.030]$ & $[-0.031,0.038]$ \Tstrut \\
$M_{\rm BC}$ [GeV$/c^2$]     & $[1.860, 1.874]$ & $-$ \Tstrut \\
M$_{\rm miss}^2$ [GeV$^2/c^4$] & $-$ & $[0.203, 0.293]$ \Tstrut \Bstrut \\
\hline
\end{tabular}
\caption{Signal $\Delta E,~M_{\rm BC}$ and M$_{\rm miss}^2$ selection bounds in data.}
\label{tab:sig_var}
\end{table}

\begin{figure}[tbp]
    \centering
    \includegraphics[scale=0.4]{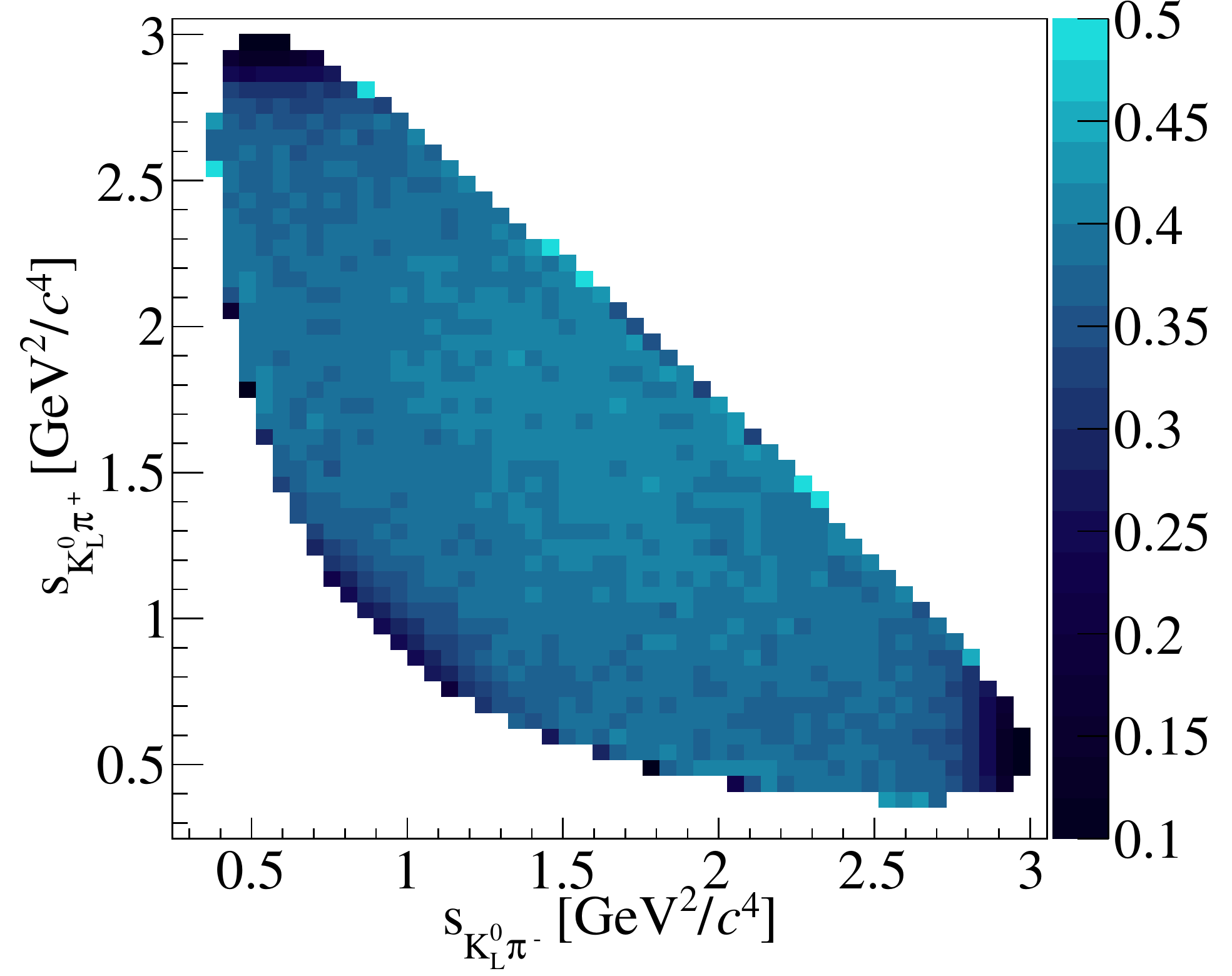}
    \caption{The DT efficiency profile for \klpipi\ signal mode tagged with the $K\pi$ candidate.}
    \label{fig:effi_prof}
\end{figure}

Using signal simulation samples for \kspipi\ and \klpipi\ modes, average double-tag efficiencies over the DP for each tag mode are calculated and are given in table~\ref{tab:DT_effi}. Figure~\ref{fig:effi_prof} shows the efficiency profile over the \klpipi\ DP for the $K^+\pi^-$ tag mode as an example. The low momentum of pions at the edges of the DP cause these regions to have reduced efficiency.

\begin{table}[tbp]
\centering
\begin{tabular}{|c|c|c|c|c|}
\hline
 & \multicolumn{2}{|c|}{ $\epsilon_{\rm avg}^{DT}$ [$\%$]} & \multicolumn{2}{|c|}{ \klpipi\ vs. tag contamination rates [$\%$]} \Tstrut \Bstrut \\
\hline
	Tag	     & $K_{\rm S}^0\pi^+\pi^-$ & $K_{\rm L}^0\pi^+\pi^-$ & Non-peaking background & \kspipi\ background  \Tstrut \Bstrut \\
		     \hline \hline
$K\pi$       &  $25.85\pm0.02$  & $35.89\pm0.02$ & $5.96\pm0.08$ & $4.79\pm0.07$  \\
$K\pi\pi\pi$ &  $12.25\pm0.02$  & $14.31\pm0.01$ & $5.84\pm0.10$ & $4.07\pm0.07$ \\
$K\pi\pi^0$  &	$13.02\pm0.01$  & $17.77\pm0.01$ & $6.48\pm0.06$ & $4.86\pm0.05$  \\ 
\hline
\end{tabular}
\caption{Average double-tag efficiencies for $K_{\rm S}^0\pi^+\pi^-$ and $K_{\rm L}^0\pi^+\pi^-$ signal modes.}
\label{tab:DT_effi}
\end{table}

Inclusive simulation samples from physics processes mentioned in section~\ref{sec:detdata} are  subjected to the same event-selection criteria as in data. Negligible background is retained with the \kspipi\ sample due to the requirement that the six-constraint kinematic fit is successful. However, the \klpipi\ selection criteria allow for about $(10-12)\%$ background contamination for the three tag modes as summarized in table~\ref{tab:DT_effi}, of which approximately $5\%$ is \kspipi, which peaks in the ${\rm M_{miss}^2}$ distribution. The remaining non-peaking background constitutes a number of hadronic and semi-leptonic modes. Modeling of these background events in the amplitude analysis is described in section~\ref{sec:ampfit}. 

\section{Amplitude analysis}
\label{sec:ampfit}

The strategy adopted is to fit the \kspipi\ sample alone to validate the fitter, then this sample is  used as a constraint in a simultaneous fit with the \klpipi\ sample to determine $\hat{\rho}_{k_{CP}}$. The signal probability distribution function (PDF) at a phase-space point $\mathbf{x}$ is expressed in terms of the total amplitude $\mathcal{A}_{\rm sig}({\bf x})$ obtained from the isobar model discussed in section~\ref{sec:theory} and the detector efficiency $\epsilon({\bf x})$:
\begin{equation}
p_{\rm sig}(\textbf{x}) = \frac{\epsilon(\textbf{x})|\mathcal{A}_{\rm sig}(\textbf{x})|^2}{\int_D \epsilon(\textbf{x})|\mathcal{A}_{\rm sig}(\textbf{x})|^2~d\textbf{x}} = \frac{\epsilon(\textbf{x})|\mathcal{A}_{\rm sig}(\textbf{x})|^2}{\mathcal{N}_{\rm sig}},
\end{equation}
where the normalization integral $\mathcal{N}_{\rm sig}$ is over the DP and is calculated using {\it Monte Carlo (MC) integration}~\cite{ref:MC_int_book}, wherein the integral can be approximated as a discrete summation of the signal PDF over a large number of phase-space points $N$ called the {\it integration sample}:
\begin{equation}
\mathcal{N}_{\rm sig} \approx \frac{1}{N}\sum_{j=0}^{N}\frac{\epsilon(\textbf{x}_j)}{q(\textbf{x}_j)}|\mathcal{A}_{\rm sig}(\textbf{x}_j)|^2.
\end{equation}
The integration sample is distributed as the PDF $q({\bf x})$ at the end of sample generation, reconstruction and selection and therefore can be related to the generator level PDF $Q({\bf x})$ as: $q({\bf x}) = \epsilon({\bf x})Q({\bf x})$. This allows cancellation of the explicit dependence on the efficiency of the normalization factor which can be written as,
\begin{equation}
\mathcal{N}_{\rm sig} = \frac{1}{N} \sum_{j=0}^{N} \frac{|\mathcal{A}_{\rm sig}(\textbf{x}_j)|^2}{Q(\textbf{x}_j)}.
\label{eq:normfacdef}
\end{equation}

The total PDF accounting for signal and background incoherently with appropriate weights is
\begin{equation}
\mathcal{P}({\bf x}) = (1 - w_{K_{\rm S}} - w_{\rm np}) p_{\rm sig}({\bf x}) + w_{K_{\rm S}} p_{K_{\rm S}}({\bf x}) + w_{\rm np} p_{\rm np}({\bf x}),
\end{equation}
where $p_{K_{\rm S}}$ and $p_{\rm np}$ are the PDFs describing \kspipi\ peaking background and the non-peaking background with weights $w_{K_{\rm S}}$ and $w_{\rm np}$ respectively. A negative log-likelihood function is constructed as
\begin{equation}
    -2{\rm log}\mathcal{L}' = -2{\rm log} \prod_{i\in {\rm data}} \mathcal{P}(\textbf{x}_i) = -2 \sum_{i\in {\rm data}} {\rm log}~ \mathcal{P}(\textbf{x}_i).
\end{equation}
Since the reconstruction efficiencies are independent of the parameters to be estimated, they can be factored out and an effective likelihood function $\mathcal{L}$ can be defined as
\begin{equation}
-2{\rm log}\mathcal{L} = -2 \sum_{i\in {\rm data}} {\rm log}~ \mathcal{P}(\textbf{x}_i) + 2\sum_{i \in {\rm data}} {\rm log}\epsilon(\textbf{x}_i); 
\end{equation}
\begin{equation}
-2 {\rm log}\mathcal{L} = -2 \sum_{i \in {\rm data}} {\rm log} \left[ (1 - w_{K_{\rm S}} - w_{\rm np}) \frac{|\mathcal{A}_{\rm sig}(\textbf{x}_i)|^2}{\mathcal{N}_{\rm sig}} + w_{K_{\rm S}} \frac{|\mathcal{A}_{K_{\rm S}}(\textbf{x}_i)|^2}{\mathcal{N}_{K_{\rm S}}} + w_{\rm np} \frac{|\mathcal{A}_{\rm np}(\textbf{x}_i)|^2}{\mathcal{N}_{\rm np}}    \right],
\label{eq:eff_LL}
\end{equation}
which can be minimized, without any efficiency function as an input, to obtain the parameters of interest ($a_r,\phi_r$) introduced in section~\ref{sec:theory}. Here, $\mathcal{N}_{K_{\rm S}}$ and $\mathcal{N}_{\rm np}$ are the normalization factors calculated using the $K_{\rm S}^0\pi^+\pi^-$ and non-peaking background model descriptions on events generated with \klpipi\ reconstruction and selection efficiency effects.

\begin{figure}
    \centering    
    \includegraphics[scale=0.7]{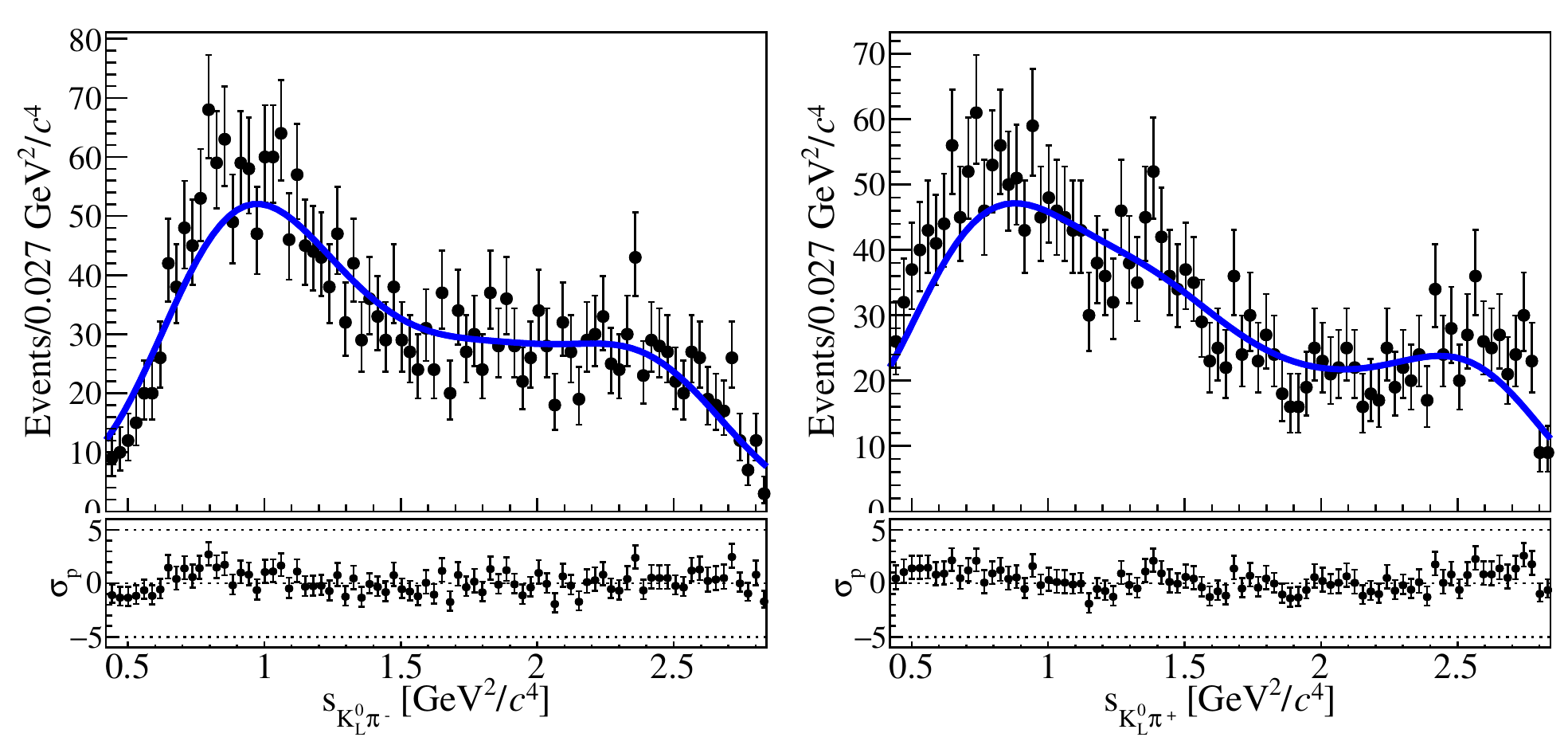}
    \caption{Distributions of $s_{K^0_{\rm L}\pi^-}$ and $s_{K^0_{\rm L}\pi^+}$ for candidates in the  ${\rm M_{miss}^2}$ sidebands (black dots) in data. The fit PDF (blue) is superimposed.}
    \label{fig:sb_pdf}    
\end{figure}

The peaking background PDF $p_{K_S}$ can be described by the same amplitude model as would be applied to the \kspipi\ signal acting as a constraint in the \klpipi\ fit. The non-peaking background PDF is described using the {\it side-bands}, {\it i.e.} regions dominated by background away from the signal, of the ${\rm M_{miss}^2}$ distribution. The width and position of the side-bands are optimized using the inclusive MC simulation sample with a $\chi^2$ statistic for maximum compatibility with the non-peaking background distribution in the signal region. The upper and lower side-band regions are defined by the limits [0.107, 0.173] GeV$^2/c^4$ and [0.323, 0.350] GeV$^2/c^4$, respectively. A two-dimensional Gaussian kernel estimator~\cite{ref:rookeys} is used to model the background distribution in the side-bands. The projections of the resultant PDF are shown in figure~\ref{fig:sb_pdf}. Any difference between the distribution of background over the DP in the sideband and the signal region is considered as a source of systematic uncertainty.

We account for two additional effects in the fit. Firstly, we correct for the fact that hadronic decays of the $D^{0}\to K^-X$, where $X$ are combinations of pions on the tag side, are not an exact flavour tag. This effect comes from a contamination by tag-side DCS decays of type $D^0 \to K^+X,~\bar{D}^0 \to K_{\rm S/L}^0\pi^+\pi^-$. An event tagged by a DCS decay will incorrectly place the signal \ksorlpipi\ at $\left(s_{K^{0}_{\rm L}\pi^-},s_{K^{0}_{\rm L}\pi^+}\right)$ rather than $\left(s_{K^{0}_{\rm L}\pi^+},s_{K^{0}_{\rm L}\pi^-}\right)$. These DCS-tagged events can be accounted for by adding their amplitudes to the nominal CF pure-flavor-tag amplitude coherently as

\begin{equation}
\mathcal{A} = \mathcal{A}_F \bar{\mathcal{A}}_{\bar{G}} - \bar{\mathcal{A}}_{\bar{F}}\mathcal{A}_{\bar{G}},
\end{equation}
where $\mathcal{A}_F(\bar{\mathcal{A}}_{\bar{F}})$ is the amplitude of the signal decay $D^0(\bar{D}^0) \to$ \kspipi\ and $\bar{\mathcal{A}}_{\bar{G}}(\mathcal{A}_{\bar{G}})$ is the amplitude of the tag mode $\bar{D}^0(D^0) \to K^+\pi^-$. The decay probabilities integrated over the allowed tag mode phase-space, at position {\bf x} on the \kspipi\ and \klpipi\ DPs, are given by

\begin{align}
|\mathcal{A}(\textbf{x})|^2_{K_{\rm S}^0\pi\pi}  &= \left( ~ |\mathcal{A}_F(\textbf{x})|^2 + (r_D^G)^2 |\bar{\mathcal{A}}_{\bar{F}}(\textbf{x})|^2 - 2r_D^G R_G \mathcal{R}e[e^{-i\delta_G}\mathcal{A}^{*}_F(\textbf{x})\bar{\mathcal{A}}_{\bar{F}}(\textbf{x})] ~   \right) , \label{eq:AAdcs_kspipi} \\
|\mathcal{A}(\textbf{x})|^2_{K_{\rm L}^0\pi\pi}  &= \left( ~ |\mathcal{A}_F(\textbf{x})|^2 + (r_D^G)^2 |\bar{\mathcal{A}}_{\bar{F}}(\textbf{x})|^2 + 2r_D^G R_G \mathcal{R}e[e^{-i\delta_G}\mathcal{A}^{*}_F(\textbf{x})\bar{\mathcal{A}}_{\bar{F}}(\textbf{x})] ~   \right) , \label{eq:AAdcs_klpipi} 
\end{align}
where the parameter $r_D^G$ is the DCS to CF amplitude ratio and $R_Ge^{-i\delta_G}$ measures the coherence between them for the tag mode $G$. The last terms in eqs.~\ref{eq:AAdcs_kspipi} and \ref{eq:AAdcs_klpipi} contain the interference between CF and DCS amplitudes, which dominates the tag-side DCS effects in the total-decay probability. The values used for the hadronic parameters for the three tag modes involved in this analysis are from the Refs. \cite{ref:lhcb_cohfac} and~\cite{ref:bes3_cohfac} and are listed in table~\ref{tab:cohfac}. 
\begin{table}[tbp]
\centering
\begin{tabular}{|c|c|c|c|}
\hline
$G$          & $r_{D}^{G}$ [$\%$] & $\delta_{G}$ [$^{\circ}$] & $R_{G}$ \Tstrut \Bstrut \\
\hline
$K\pi$       & $5.867\pm0.015$    & $190.0^{+4.2}_{-4.1}$ & 1 \Tstrut \\
$K\pi\pi\pi$ & $5.50\pm0.07$    & $161^{+28}_{-18}$      & $0.44^{+0.10}_{-0.09}$ \Tstrut \\
$K\pi\pi^0$  & $4.41\pm0.11$    & $196 \pm 11$      & $0.79\pm 0.04$ \Tstrut \\
\hline
\end{tabular}
\caption{DCS to CF amplitude ratios and coherence parameters~\cite{ref:lhcb_cohfac, ref:bes3_cohfac}.}
\label{tab:cohfac}
\end{table}

Secondly, we account for differences in the acceptance between the experimental data and the simulated events used to normalize the PDFs. Any difference is accounted for in the total PDF by scaling the normalization factor by ratios of reconstruction efficiencies in data $\epsilon^{\rm data}$ and simulated events $\epsilon^{\rm sim}$, obtained by studying BESIII control samples, for each of the final state particles in the signal mode. These scale factors denoted by $\gamma_{\epsilon}$ are defined as
\begin{equation}
\gamma_{\epsilon} = \prod_i \frac{\epsilon^{\rm data}_i(p_i)}{\epsilon^{\rm sim}_i(p_i)},
\end{equation}
where $p_i$ is momentum of the decay particle $i \in \lbrace K_{\rm S/L}^0, \pi^+, \pi^- \rbrace$. The contributions to the correction for pions arise from both PID and tracking efficiency differences. The $\gamma_{\epsilon}$ factor variations with momentum are parametrized using various polynomial and exponential functions and the net correction in the normalisation factor defined in eq.~\ref{eq:normfacdef} appears as,
\begin{equation}
\mathcal{N} = \frac{1}{N} \sum_{j=0}^N \gamma_{\epsilon}(p_j^{K^0}, p_j^{\pi^+}, p_j^{\pi^-})~|\mathcal{A}(\textbf{x}_j)|^2,
\end{equation}   where, $p_j^{K^0}, p_j^{\pi^+}$ and $p_j^{\pi^-}$ are the momenta of $K_{\rm S/L},~\pi^+$ and $\pi^-$ at point $\textbf{x}$ on the DP, respectively.

Previous \kspipi\ amplitude-model analysis based on Belle and \textsc{BaBar} data suggests eleven Breit-Wigner $P$ and $D$ wave resonances, two $K\pi$ S-wave and a broad $\pi\pi$ S-wave contribution to the three-body decay \cite{ref:bellebabar18}. However, the BESIII data sample is two orders of magnitude smaller than the combined Belle and \textsc{BaBar} data set, which means there is no sensitivity to some of the components. We assign significance to each of the contributions as p-values from $\chi^2$ distributions, approximated using Wilk's theorem~\cite{ref:wilks}, from the ratios of generalized log-likelihood values, calculated with and without the resonant contribution under test. The significance values for both the scenarios against each of the test resonance are given in table~\ref{tab:signi_vals}. The resonance $\rho(770)$ is considered as the reference relative to which all other resonance amplitudes are measured. In a standalone \kspipi\ amplitude fit, the resonances $K^{*}(1680)^-,~K^{*}(892)^+,~K_{2}^{*}(1430)^+,~K^{*}(1410)^+$ and $K_0^{*}(1430)^+$ are observed to be statistically insignificant, whereas upon including the \klpipi\ mode in the fit, only the DCS $K_0^{*}(1430)^+$ has a significance below $3\sigma$.

\begin{table}[tbp]
\centering
\begin{tabular}{|c|c|c|}
\hline
Resonance & [$K_{\rm S}^0\pi^+\pi^-$] & [$K_{\rm L}^0\pi^+\pi^-,$ $K_{\rm S}^0\pi^+\pi^-$]  \Tstrut \Bstrut  \\
\hline \hline
$K^0\rho(770)$   		 & $-$     & $-$ \Tstrut \\
$K^0\omega(782)$ 		 & $5.1$   & $8.8$ \Tstrut \\
$K^0f_{2}(1270)$ 		 & $7.3$   & $11.9$ \Tstrut \\ 
$K^0\rho(1450)$  		 & $4.2$   & $12.1$ \Tstrut \\
$K^{*}(892)^-\pi^+$ 	 & $86.3$  & $168.9$ \Tstrut \\
$K_2^{*}(1430)^-\pi^+$   & $12.4$  & $19.1$ \Tstrut \\
$K^{*}(1680)^-\pi^+$  	 & $1.7$   & $5.7$ \Tstrut \\
$K^{*}(1410)^-\pi^+$  	 & $4.2$   & $3.7$ \Tstrut \\
$K^{*}(892)^+\pi^-$   	 & $0.5$   & $17.0$ \Tstrut \\
$K_{2}^{*}(1430)^+\pi^-$ & $2.0$   & $4.3$ \Tstrut \\
$K^{*}(1410)^+\pi^-$     & $1.1$   & $2.9$ \Tstrut \\
$K_0^{*}(1430)^-\pi^+$   & $20.1$  & $37.1$ \Tstrut \\
$K_0^{*}(1430)^+\pi^-$   & $1.1$   & $2.0$ \Tstrut \\
\hline
$\pi^+\pi^-$ S-wave      & $32.1$  & $52.8$ \Tstrut \\
\hline
\end{tabular}
\caption{Significance values of resonant components for standalone $K_{\rm S}^0\pi^+\pi^-$ and simultaneous $K_{\rm L}^0\pi^+\pi^-,$ $K_{\rm S}^0\pi^+\pi^-$ model fits.}
\label{tab:signi_vals}
\end{table}

The first validation of the fitter is carried out by generating an ensemble of 350 simulated \kspipi\ and \klpipi\ event samples based on a fitted model as is described in section~\ref{sec:fitres} ahead, which are fit to compare the measured and generated model parameters. Each generated sample has the same size as the data sample. The \klpipi\ samples contain DCS as well as background contamination in the same proportion as data for all the three tag modes. The $K$-matrix parameters are global parameters and their values are known from the large Belle-\textsc{BaBar} data sample~\cite{ref:bellebabar18}. Therefore, the $K$-matrix description is kept fixed in all the fits. Distributions of the difference between input and output model parameters divided by their uncertainty are produced and found to be consistent with a normal distribution.

\begin{figure}
    \centering
    \includegraphics[scale=0.75]{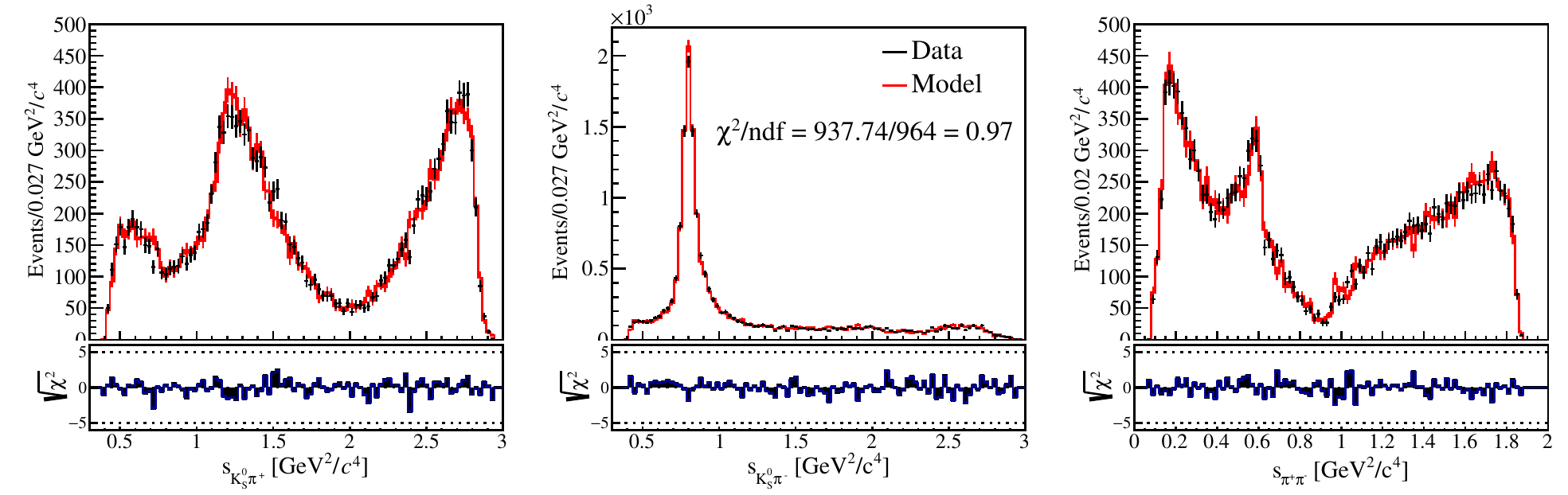}
    \caption{DP projections of \kspipi\ data event distribution and model prediction from a standalone \kspipi\ fit.}
    \label{fig:ksonly_fitproj}
\end{figure}

The second validation is to compare to the results of Ref.~\cite{ref:bellebabar18}. Therefore, we perform the analysis on a \kspipi\ candidates only. The fit model is compared against data as the DP projections and is shown in figure~\ref{fig:ksonly_fitproj}. The goodness of fit is measured with a reduced $\chi^2$ statistic and is observed to be equal to 0.97 for the toy model fit. To further validate the fit, we determine the fit fractions of our model. In contrast to the amplitude formalism dependent fit parameters $a_r$ and $\phi_r$, the fractional contribution of each component to the total decay probability, called {\it fit fraction}, is expected to be a consistent global physics parameter for a particular multibody decay. The functional form of the fit fraction for a resonant contribution $R$ is given by,
\begin{equation}
FF_R = \frac{\int {\rm d}{\bf x}~ |\mathcal{A}_R(\textbf{x})|^2 }{\int {\rm d}{\bf x}~ |\sum_{r}\mathcal{A}_r(\textbf{x})|^2}.
\end{equation}
An aggregate of the fit fractions away from $100\%$ suggests interference effects. The predicted fit fractions of various components show reasonable agreement with the Belle-\textsc{BaBar} values as also shown in table~\ref{tab:ffr_kspipi_SA}. 

\begin{table}[ht!]
\centering
\begin{tabular}{|c|c|c|}
\hline
Resonance & Belle-\textsc{BaBar} $FF_R$ [$\%$]  & Predicted $FF_R$ [$\%$]  \Tstrut \\[0.5ex]
\hline \hline
$K_{S}^{0}\rho(770)^{0}$     & $20.4$  & $17.2_{-1.9}^{+1.8}$ \Tstrut  \\
$K_{S}^{0}\omega(782)$       & $0.5$   & $0.3_{-0.1}^{+0.3}$ \Tstrut \\
$K_{S}^{0}f_{2}(1270)$       & $0.8$   & $0.7_{-0.8}^{+0.7}$ \Tstrut \\ 
$K_{S}^{0}\rho(1450)^{0}$    & $0.6$   & $0.8_{-1.2}^{+1.1}$ \Tstrut \\
$K^{*}(892)^{-}\pi^{+}$      & $59.9$  & $58.9_{-3.6}^{+6.7}$ \Tstrut \\
$K^{*}_{2}(1430)^{-}\pi^{+}$ & $1.3$   & $2.1_{-0.4}^{+0.2}$ \Tstrut \\ 
$K^{*}(1680)^{-}\pi^{+}$     & $0.5$   & $0.52 \pm 0.02$ \Tstrut \\
$K^{*}(1410)^{-}\pi^{+}$     & $0.1$   & $0.020 \pm 0.001$ \Tstrut \\
$K^{*}(892)^{+}\pi^{-}$      & $0.6$   & $0.54 \pm 0.01$ \Tstrut \\ 
$K^{*}_{2}(1430)^{+}\pi^{-}$ & $<0.1$   & $0.011 \pm 0.001$ \Tstrut \\
$K^{*}(1410)^{+}\pi^{-}$     & $<0.1$   & $0.0304 \pm 0.0001$ \Tstrut \\ 
\hline
$\pi^{+}\pi^{-}$ $S$-wave    & $10.0$ & $8.2_{-0.5}^{+1.0}$ \Tstrut \\  
$K^{*}_{0}(1430)^{-}\pi^{+}$ & $7.0$ & $8.2_{-2.2}^{+1.0}$ \Tstrut	  \\
$K^{*}_{0}(1430)^{+}\pi^{-}$ & $<0.1$ & $0.0084 \pm 0.0002$	\Tstrut		 \\
\hline
Total & 101.6  & $100.2_{-10.7}^{+12.8}$ \Tstrut \\ [0.5ex]
\hline
\end{tabular}
\caption{Comparison of predicted fit fractions from the standalone \kspipi\ model with Belle-\textsc{BaBar} results.}
\label{tab:ffr_kspipi_SA}
\end{table}

\section{Results}
\label{sec:fitres}

The results are obtained from a simultaneous fit to the combined sample of \kspipi\ and \klpipi\ data candidates. The fit minimizes the negative log-likelihood function composed of the PDFs given in eqs.~\ref{eq:AAdcs_kspipi} and \ref{eq:AAdcs_klpipi}, which include all the statistically significant (greater than $3\sigma$) components listed in table~\ref{tab:signi_vals}, as well as incorporating the multiplicative factors related to the U-spin breaking parameters for the $\mathit{CP}$ resonances, described in section~\ref{sec:theory}. The peaking background fraction is fixed to the values obtained from simulated events, whereas the non-peaking background fractions are constrained with an additional $\chi^2$ term in the likelihood function, composed of weighted difference between MC simulated values and the values to be estimated. Uncertainties associated with both assumptions are considered as systematic uncertainties. The amplitude and U-spin breaking parameters are presented in tables~\ref{tab:arphir_vals} and~\ref{tab:uspin_final}, respectively. The modulus of the $\hat{\rho}$ parameters notably lie in a wide range of values from 0.4 for the $\pi\pi$ S-wave to 12.1 for the $\rho(1450)$ resonance and the phases are in general measured to be away from $0^{\circ}$ for all the $\mathit{CP}$ resonances. 
\begin{table}[tbp]
\centering
\begin{tabular}{|c|c|c|}
\hline
Resonance & $a_r$ & $\phi_r$ [$ ^{\circ}$] \Tstrut \Bstrut  \\
\hline
$\rho(770)$           & 1.0                &            0.0  \\
$\omega(782)$         & $0.0388\pm 0.0031$ &     $ 105.1\pm 5.3$ \Tstrut \\
$f_{2}(1270)$         & $1.30\pm 0.13$     &     $-41.2\pm 4.4$ \Tstrut \\
$\rho(1450)$          & $1.69\pm 0.40$     &     $109.9\pm 8.2$ \Tstrut \\
$K^{*}(892)^{-}$      & $1.84\pm 0.03$     &     $138.3\pm 1.2$ \Tstrut \\
$K_{2}^{*}(1430)^{-}$ & $1.50\pm 0.05$     &     $-48.0\pm 2.1$ \Tstrut \\
$K^{*}(1680)^{-}$     & $2.32\pm 0.28$     &     $-191.5\pm 9.4$ \Tstrut \\
$K^{*}(1410)^{-}$     & $0.48\pm 0.07$     &     $-143.0\pm 7.6$ \Tstrut \\
$K^{*}(892)^{+}$      & $0.16\pm 0.01$     &     $-36.2\pm 3.6$ \Tstrut \\
$K_{2}^{*}(1430)^{+}$ & $0.25\pm 0.05$     &     $-86.6\pm 10.9$ \Tstrut \\
$K^{*}(1410)^{+}$     & $0.23\pm 0.05$     &     $68.5\pm 12.9$ \Tstrut \\
$K_{0}^{*}(1430)^{-}$ & $2.39\pm 0.06$     &     $97.7\pm 1.4$ \Tstrut  \\

\hline

\end{tabular}
\caption{Amplitude parameters ($a_r,\phi_r$) predicted from the simultaneous fit.}
\label{tab:arphir_vals}
\end{table}

\begin{figure}[tbp]
    \centering
    \includegraphics[scale=0.75]{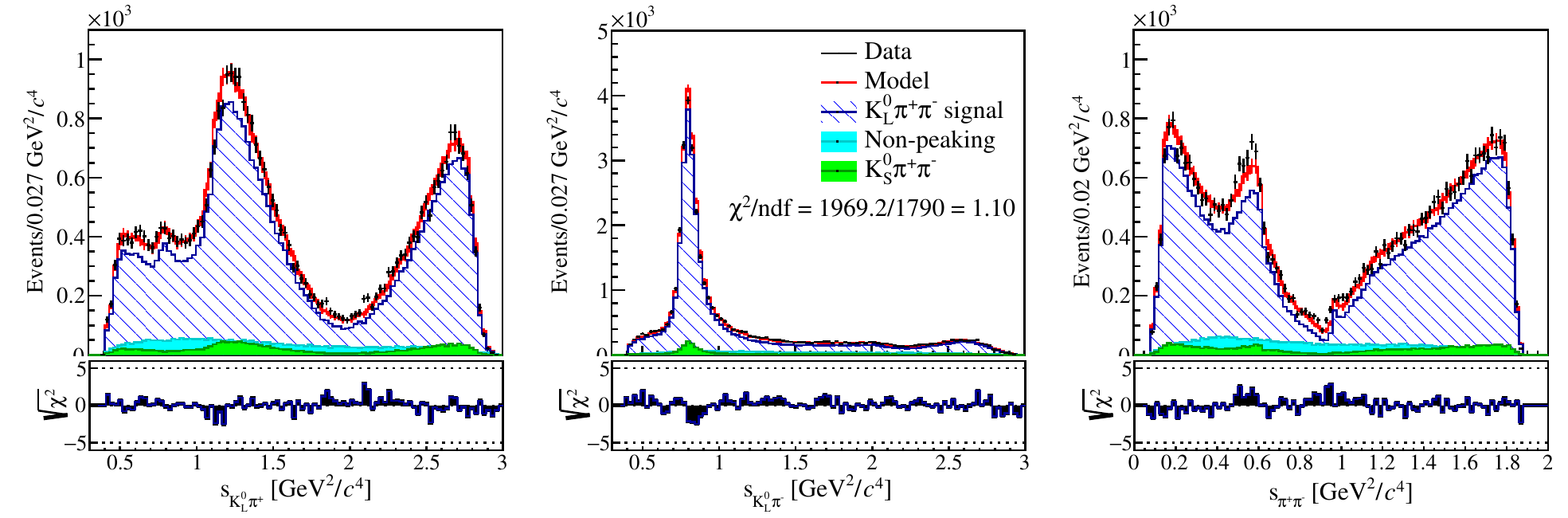}
    \caption{Model and data DP projections of \klpipi\ signal mode.}
    \label{fig:klpipi_fitproj}
\end{figure}

\begin{figure}[h!]
    \centering
    \includegraphics[scale=0.75]{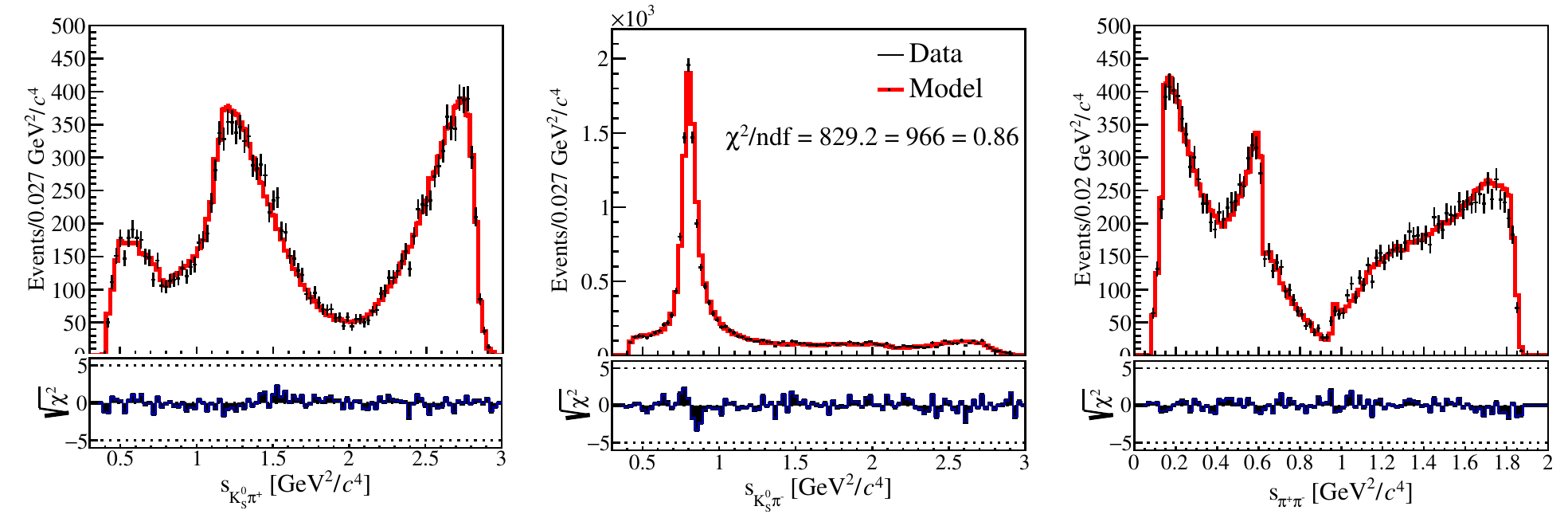}
    \caption{Model and data DP projections of \kspipi\ signal mode.}
    \label{fig:kspipi_fitproj}
\end{figure}

\begin{table}[tbp]
\centering
\small
\begin{tabular}{|c|c|c|c|}
\hline
Resonance & $|\hat{\rho}|$ & arg($\hat{\rho}$) [$^{\circ}$] & $|1-2{\rm tan}^2\theta_C\hat{\rho}|^2$ \Tstrut \Bstrut \\
\hline
$\rho(770)$     & $1.93\pm0.27\pm0.42$ & $-90.6\pm5.8\pm 7.6$    & $1.05\pm0.04\pm0.06$ \Tstrut \\
$\omega(782)$   & $6.13\pm0.75\pm0.53$ & $~2.2\pm7.0\pm 4.8$     & $0.12\pm0.05\pm0.04$ \Tstrut \\
$f_{2}(1270)$   & $3.75\pm0.90\pm0.81$ & $-56.5\pm16.8\pm 12.9$  & $0.72\pm0.20\pm0.15$ \Tstrut \\
$\rho(1450)$    & $12.12\pm2.92\pm1.88$ & $~78.4\pm14.4\pm 15.6$ & $2.19\pm0.95\pm0.83$ \Tstrut \\
$\pi\pi$ S-wave & $0.37\pm0.21\pm0.37$ & $-164.4\pm15.7\pm 13.4$ & $1.08\pm0.04\pm0.08$ \Tstrut \\
\hline
\end{tabular}
\caption{Measured U-spin breaking parameters. The first uncertainty is statistical and the second is total systematic.}
\label{tab:uspin_final}
\end{table}
DP projections of the predicted model compared to data for \klpipi\ and \kspipi\ modes are given in figures~\ref{fig:klpipi_fitproj} and \ref{fig:kspipi_fitproj}, respectively. The reduced $\chi^2$, after ensuring statistical significance in each 2D phase-space bin through combining adjacent bins, is found to be $1969.2/1790 = 1.10$ for \klpipi\ and $829.2/966 = 0.86$ for \kspipi\ mode, suggesting that the model describes the data reasonably well. Small deviations are observed in the $\rho(770)-\omega(782)$ interference region. The DCS interference in a \klpipi\ model has overall constructive effects as opposed to an overall destructive interference in a \kspipi\ model. This effect results in the lower total fit fraction and in the partial-fit fractions of some CF components, such as $K^{*}(892)^-$ and $K_0^{*}(1430)^-$ for a \klpipi\ model as compared to \kspipi. The $CP$-resonance fit fractions are additionally affected by the U-spin breaking parameter phases, of which the $\omega(782)$ resonance is a clear example. Fit fraction for both the modes are given in table~\ref{tab:k0lspipi_ffr}.

\begin{figure}
\begin{subfigure}[t]{0.48\textwidth}
\includegraphics[scale=0.35]{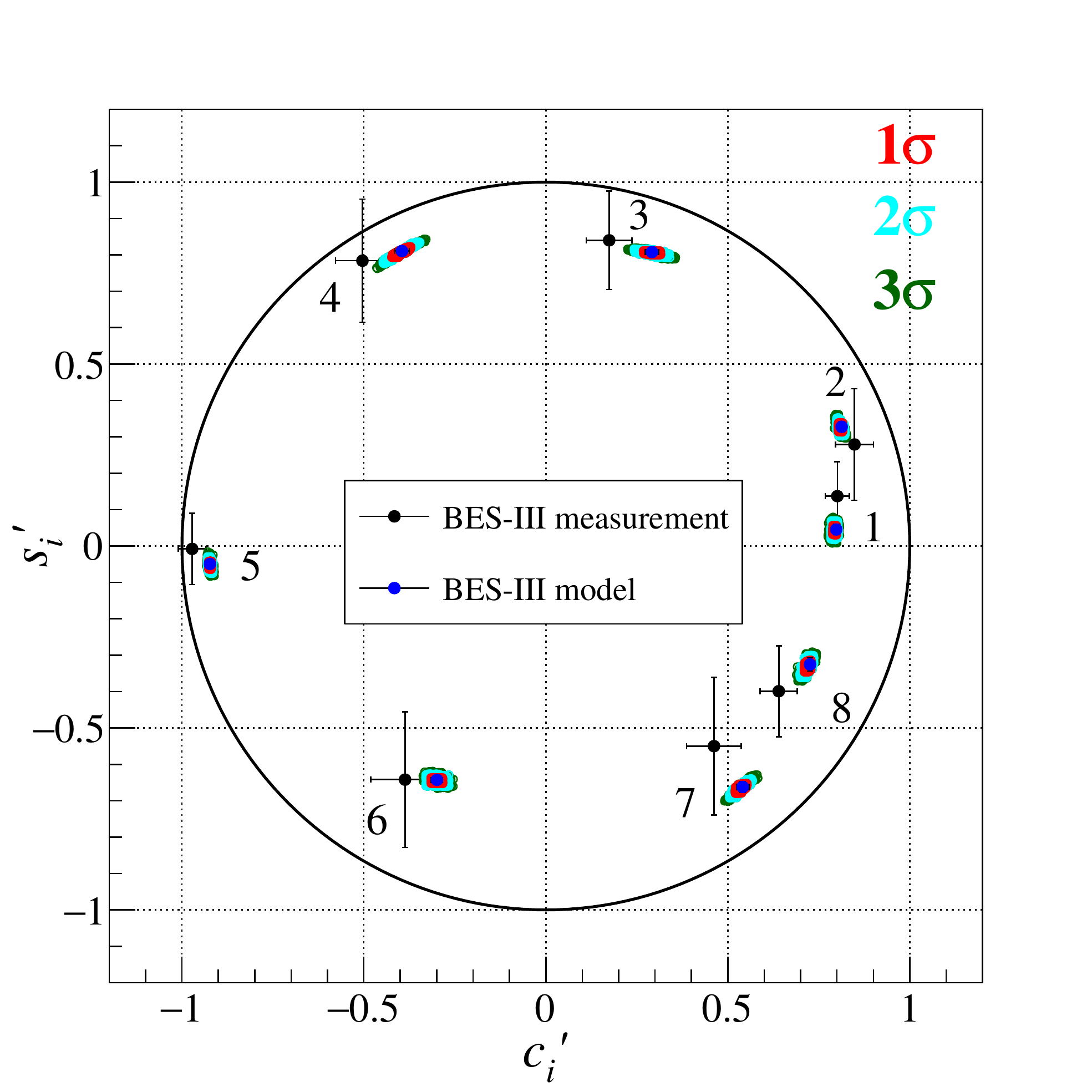}
\end{subfigure}
\hfill
\begin{subfigure}[t]{0.48\textwidth}
\includegraphics[scale=0.35]{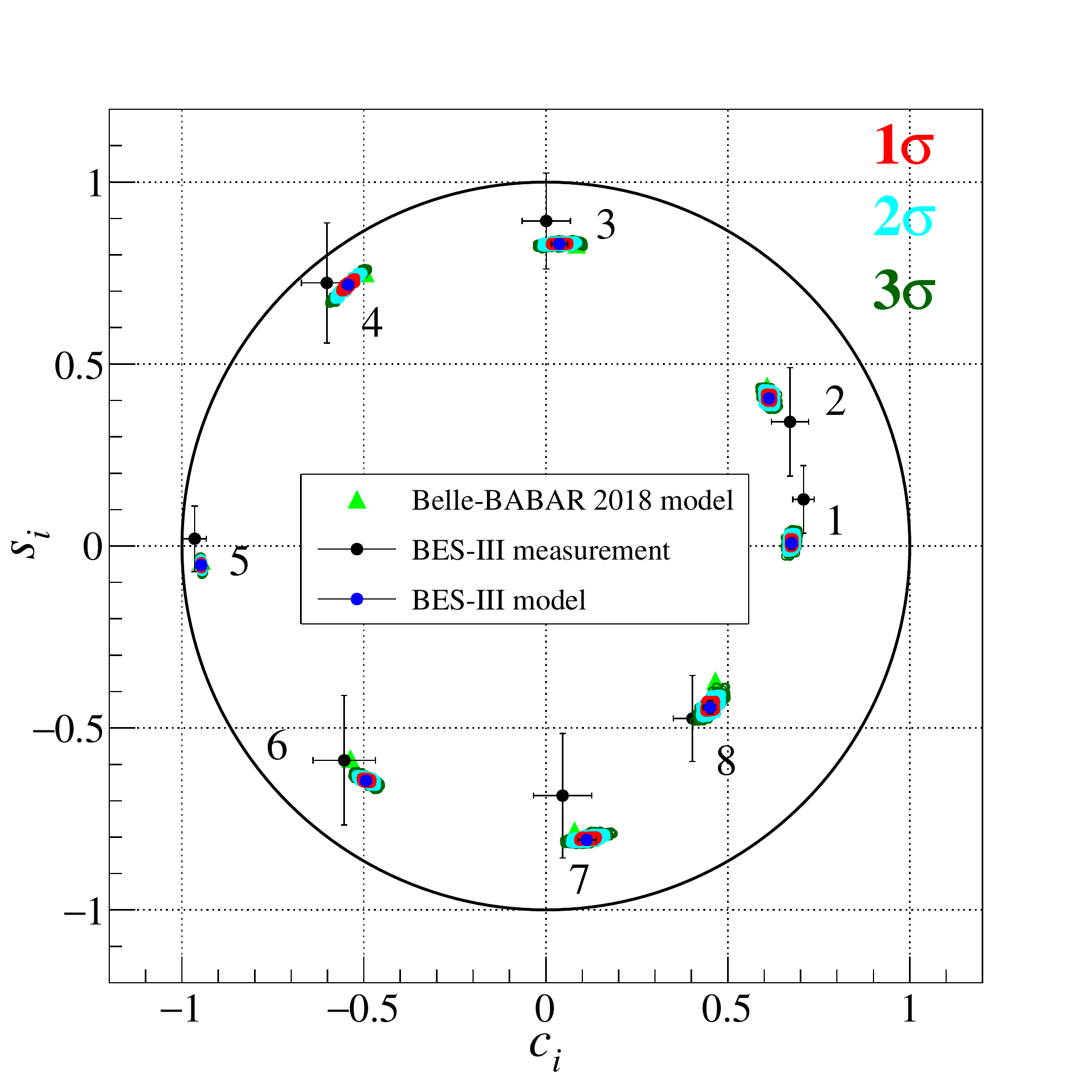}
\end{subfigure}
\caption{({\it Left}) Strong-phase parameters $c_i'$ and $s_i'$ predicted using the $K_{\rm L}^0\pi^+\pi^-$ model from this analysis, compared against the model-independent BESIII measurements~\cite{ref:bellebabar18}. ({\it Right}) Strong-phase parameters $c_i$ and $s_i$ predicted using the $K_{\rm S}^0\pi^+\pi^-$ model from this analysis, compared against the model-independent BESIII measurements and the Belle model-predicted values.}
\label{fig:cisicipsip}
\end{figure}

Model-predicted strong-phase parameter values for $K_{\rm L}^0\pi^+\pi^-$ and $K_{\rm S}^0\pi^+\pi^-$, in equal-$\Delta \delta_D$ binning, are presented in figure~\ref{fig:cisicipsip}, along with statistical error coverage up to three standard deviations. Model-independent and Belle-\textsc{BaBar} \kspipi\ model \cite{ref:bellebabar18} predicted values are also shown for comparison. These show good agreement with the model-independent measurements, with the reduced $\chi^2$ values of 1.12 for ($c_i,s_i$) and 0.21 for ($c_i',s_i'$), weighted by both statistical and systematic uncertainties.

As part of robustness tests on the amplitude model, various crosschecks on the amplitude fit are performed. The nominal isobar model for $\rho(770)$ and $\omega(782)$ is replaced with a $\rho-\omega$ mixing model with $\rho(770)$ parametrized by a Gounaris-Sakurai lineshape~\cite{ref:GS} instead of the relativistic Breit-Wigner. The fit yields worse agreement with data at the $\rho(770)$ peak and in the $\rho(770)-\omega(782)$ interference region and therefore the nominal isobar model is preferred.

Separate fits on data subdivided by tag-mode are performed. The resultant U-spin breaking parameters from all the three optimizations agree with the main fit results within uncertainties and are provided in appendix~\ref{appensec:septagfits}. An additional test is performed by modeling the \klpipi\ data independently without using the \kspipi\ data as a constraint. The procedure and the results are provided in appendix~\ref{appensec:klonly}.   

\begin{table}[tbp]
\centering
\begin{tabular}{|c|c|c|}
\hline
Resonance & $K_{\rm L}^0\pi^+\pi^-$ $FF_R$ [$\%$] & $K_{\rm S}^0\pi^+\pi^-$ $FF_R$ [$\%$] \Tstrut \Bstrut \\
\hline
$\rho(770)$           & $18.16^{+0.53}_{-0.45} \pm 2.50$ & $18.90\pm 0.42 \pm 2.12$ \Tstrut \\
$\omega(782)$         & $0.06_{-0.02}^{+0.03} \pm 0.04$  & $0.54 \pm 0.09 \pm 0.14$ \Tstrut \\
$f_{2}(1270)$         & $0.40 \pm 0.08 \pm 0.37$         & $0.61_{-0.11}^{+0.13} \pm 0.29$ \Tstrut \\
$\rho(1450)$          & $0.42 \pm 0.08 \pm 0.53$         & $0.21 \pm 0.10 \pm 0.40$ \Tstrut \\
$K^{*}(892)^{-}$      & $56.98^{+0.58}_{-0.56} \pm 3.10$ & $62.18_{-0.59}^{+0.55} \pm 2.58$ \Tstrut \\
$K_{2}^{*}(1430)^{-}$ & $1.64^{+0.10}_{-0.09} \pm 0.48$  & $1.79 \pm 0.09 \pm 0.47$ \Tstrut \\
$K^{*}(1680)^{-}$     & $0.25^{+0.06}_{-0.05} \pm 0.68$  & $0.27 \pm 0.06 \pm 0.63$ \Tstrut \\
$K^{*}(1410)^{-}$     & $0.19 \pm 0.06 \pm 0.46$  		& $0.21 \pm 0.06 \pm 0.19$ \Tstrut \\
$K^{*}(892)^{+}$      & $0.45 \pm 0.05 \pm 0.14$  		& $0.49 \pm 0.05 \pm 0.35$ \Tstrut \\
$K_{2}^{*}(1430)^{+}$ & $0.05 \pm 0.02 \pm 0.04$  		& $0.05 \pm 0.02 \pm 0.03$ \Tstrut \\
$K^{*}(1410)^{+}$     & $0.04 \pm 0.02 \pm 0.03$         & $0.05 \pm 0.02 \pm 0.02$ \Tstrut \\ 
$K_{0}^{*}(1430)^{-}$ & $6.84^{+0.24}_{-0.25} \pm 1.84$  & $7.47 \pm 0.26 \pm 1.55$ \Tstrut \Bstrut \\
\hline
$\pi\pi$ S-wave       & $10.12_{-0.33}^{+0.32} \pm 0.96$ & $10.24 \pm 0.23 \pm 1.62$ \Tstrut \Bstrut \\
\hline \hline
Total              & $95.59^{+2.16}_{-2.07} \pm 11.17$ & $103.02_{-2.10}^{+2.11} \pm 10.39$ \Tstrut \Bstrut \\
\hline
\end{tabular}
\caption{$D^0 \to K_{\rm L,S}^0\pi^+\pi^-$ fit fractions from the simultaneous amplitude fit.}
\label{tab:k0lspipi_ffr}
\end{table}

\section{Systematic uncertainty}
\label{ssec:syst}
Several sources of systematic uncertainty are explored. Parameters that are kept fixed in the fit are the dominant contribution; the next largest contributions come from experimental effects such as acceptance. 

The fixed parameters include masses and widths of the resonant states, the $K-$matrix coupling parameters $\beta_{\alpha}$ and $f^{\rm prod}_{1i}$ for four poles and four channels, LASS resonant and non-resonant relative magnitudes and phases, the effective radii, and the tag mode DCS to CF ratios and coherence factors. Uncertainties due to these are calculated by performing repeated simultaneous \kspipi\ and \klpipi\ data fits on smeared values of the concerned fixed parameter within its uncertainty. Fixed central values and uncertainties on masses, widths and meson radii are taken from the PDG, on $K-$matrix and LASS parameters from the results of Belle-\textsc{BaBar} analysis \cite{ref:bellebabar18}, and the latest LHCb~\cite{ref:lhcb_cohfac} and BESIII~\cite{ref:bes3_cohfac} results are used for the hadronic decay parameters of the tag modes. As the uncertainties on the meson radii are actually ignorance on their values, as there have been no experimental measurements, their smearing is uniform. Central values used for effective radii for the resonances is 1.5 GeV$^{-1}$ and for the $D$ meson is 5 GeV$^{-1}$, and the ignorance is valued at $\pm 1$~GeV$^{-1}$ for both. Gaussian smearing is employed for the rest of the fixed parameters.

The effects of uncertainty associated with the difference in acceptance of data and simulated events parametrized by the $\gamma_{\epsilon}$ factor is studied. Binned uncertainties for pion and kaon PID and tracking efficiency, as well as $K_{\rm S,L}^0$ reconstruction efficiencies are taken from the control sample studies mentioned in section~\ref{sec:ampfit}. We perform data fits varying the values of $\gamma_{\epsilon}$ by one standard deviation and the variations in the fit parameters are used to assign uncertainties from this source. 

Minor contributions to the systematic uncertainty are associated with the peaking-background fraction ($w_{K_S}$) and the non-peaking background description on the phase-space. Uncertainty on the peaking background fraction comes from the size of the simulated sample, which is less than 0.1$\%$ absolute and \ks\ $\to$ \kl\ mis-ID rate, calculated by reconstructing $K_{\rm S}^0 \to \pi^0\pi^0$ in data using a $J/\psi$ control sample, is about 4.5$\%$ of $w_{K_S}$. To calculate the uncertainty associated with the modeling of the non-peaking background on the phase-space, pseudo-experiment \klpipi\ data samples are generated wherein the non-peaking background component is modeled from the ${\rm M_{miss}^2}$ signal region in the simulated sample. Amplitude fits are performed on these pseudo-experiment samples with non-peaking background descriptions based first on the actual signal region background events and then on the sideband event distributions. Any departure observed in the output values of these two cases is assigned as the systematic uncertainty from this source and is of the order of a percent for the U-spin breaking parameters.   

The total systematic uncertainty is obtained by adding all the sources in quadrature and is found to be of a similar size to the statistical uncertainties for the U-spin breaking parameters. The fit fractions carry significantly larger systematic uncertainty as compared to statistical. The systematic uncertainties on U-spin breaking parameters, fit fractions and strong-phase parameters due to each source are given in appendix~\ref{appensec:systbreakup}.

\section{CP content and BF ratio}
\label{sec:cpcontent}
Functional form of the decay amplitudes on the phase-space of the \kspipi\ and \klpipi\ final states can be exploited to calculate the $CP$ even fraction of both the states and the ratio of their branching fractions. The $CP$ even fraction of a multi-particle state like $K_{\rm L}^0\pi^+\pi^-$ is defined as~\cite{ref:cpcontent},
\begin{equation}
    F_{CP} = \frac{M_+}{M_++M_-} = \frac{1}{2} - {\rm C}\sqrt{F\bar{F}}, 
    \label{eq:cpevenfrac}
\end{equation}
where $M_{+(-)}$ is the decay probability of $D^0$ into \klpipi\ in a $CP$ even(odd) state and C is the weighted average of cosine of the associated strong-phase difference over the entire phase-space and an unbinned version of eq.~\ref{eq:cisidef}. $F$ and $\bar{F}$ are the fractions of flavor-tagged \dtoklpipi\ and \dbartoklpipi\ yields, respectively. A similar treatment can be applied to \kspipi\ state by adding the C$\sqrt{F\bar{F}}$ factor to 0.5. The model predicted $F_{CP}$ value for \klpipi\ state is found to be $35.3\pm0.6_{\rm (stat.)}\pm1.4_{\rm (syst.)}\%$ and for \kspipi\ $55.6\pm0.6_{\rm (stat.)}\pm1.2_{\rm (syst.)}\%$. These results agree with the values $35.2\%$ for \klpipi\ and $55.2\%$ for \kspipi\ calculated from Ref.~\cite{ref:bes3prd_cisi} with measured $CP$ yields in a model-independent approach. This suggests that the state \klpipi\ is significantly $CP$ odd in contrast to the \kspipi\ state, which is approximately $CP$ neutral.

Finally, the ratio of branching fractions of \dtoklpipi\ and \dtokspipi\ decay modes from a model can be calculated by dividing the respective total decay probabilities on a common phase-space distribution, and is evaluated as $1.091\pm0.012_{({\rm stat.})}\pm0.032_{({\rm syst.})}$. This is in good agreement with the corresponding number from a model-independent analysis, which is estimated to be $1.105\pm0.012_{({\rm stat.})}\pm0.015_{({\rm syst.})}$ using inputs from Refs.~\cite{ref:bes3prd_cisi, ref:bes3prl_cisi}.

\section{Conclusion}
\label{sec:conc}
Using quantum-correlated $D\bar{D}$ pairs, the first data-driven determination of the U-spin breaking parameters associated with the decay \dtoklpipi is reported. The measured values for all the $CP$-resonant modes show significant deviations from the nominally assumed value of unity. For all resonant modes, we place tighter bounds on [$|\hat{\rho}|$, arg($\hat{\rho}$)] than the previously assumed values of [0.5, 360$^{\circ}$]~\cite{ref:bes3prd_cisi}. Consequently, U-spin breaking effects manifest themselves as a considerable asymmetry between \kspipi\ and \klpipi\ fit fractions of $CP$ resonant modes like $\omega(782)$. 

\begin{figure}[tbp]
    \centering
    \includegraphics[scale=0.5]{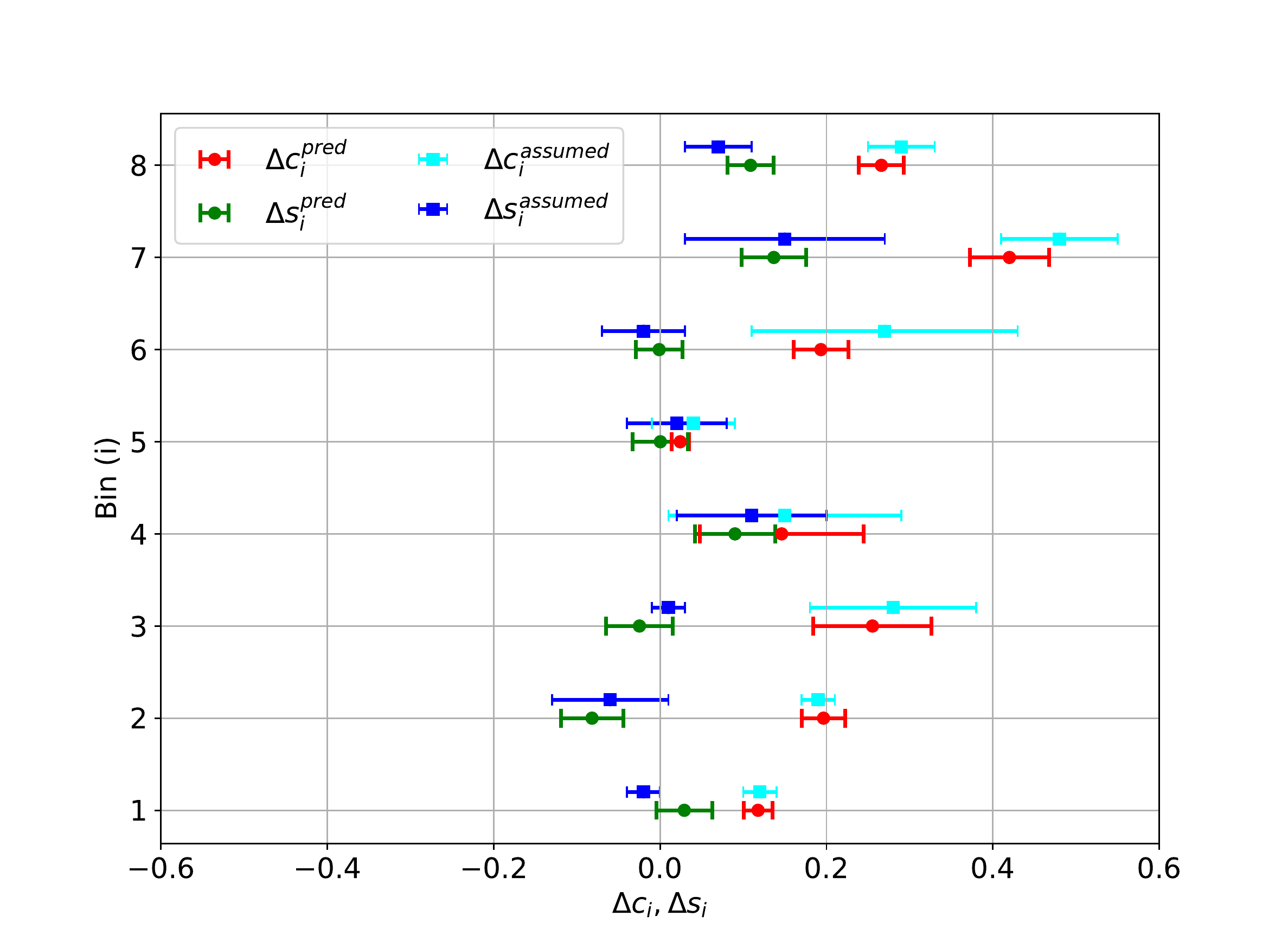}
    \caption{Model-predicted strong-phase parameter differences compared with the assumed values in the model-independent analysis from BESIII~\cite{ref:bes3prd_cisi}.}
    \label{fig:Deltacisi}
\end{figure}

Furthermore, model-predicted strong-phase parameter differences between \kspipi\ and \klpipi\ ($\Delta c_i^{\rm pred}$ and $\Delta s_i^{\rm pred}$) are calculated and are presented in figure~\ref{fig:Deltacisi} with a comparison with the values used in the model-independent strong-phase measurement from BESIII~\cite{ref:bes3prd_cisi}. The values are also given in table~\ref{tab:DciDsi} in appendix~\ref{appen:DciDsi}.  The uncertainties on the model-predictions from this analysis include both statistical and systematic uncertainties. The uncertainties on the assumed values in the model-independent analysis are determined by smearing the $r$ values in a Gaussian distribution with $0.5$ standard deviation about a mean of unity and $\delta$ values uniformly in the full range. Additionally, difference between the \textsc{BaBar} 2005 and Belle 2010 \kspipi\ models are also included, both of which consider the same $CP$ intermediate resonances in the decay described with Breit-Wigner functions for both $P$ and $S$ waves, under SU(3) flavour symmetry assumption. This results in smaller uncertainties on the assumed values in bins 1 and 3, which contain major contributions from the $\rho(770)$ and $\pi\pi$ S-wave intermediate states. The predicted values in these bins carry uncertainties from the U-spin breaking parameters. The uncertainties in the rest of the bins are reduced as compared with the assumed values, which will result in reducing the systematic uncertainty related to the U-spin assumption in future determinations of $(c_i,s_i)$.

\clearpage
\appendix

\section{Angular dependence of Breit-Wigner amplitude in \texorpdfstring{$D\to PV$}{dtopv} decays}
\label{appensec:BW_spin}
For reactions of type $D\to (ab)_rc$ and with spin transfer $0 \to j + l,$ where $l$ ($=j$) is the three-dimensional orbital angular momentum of the spectator particle, the angular part of the amplitude reduces to the expressions given in eqs.~\ref{Eq:zl0}, \ref{Eq:zl1} and \ref{Eq:zl2}.  

\begin{align}
\mathcal{Z}_{L=0} &= 1, \label{Eq:zl0}\\
\mathcal{Z}_{L=1} &= M_{bc}^2 - M_{ac}^2 - \frac{(M_D^2-M_c^2)(M_b^2-M_a^2)}{M_{ab}^2}, \label{Eq:zl1}\\
\mathcal{Z}_{L=2} &= a_1^2 - \frac{1}{3}a_2^2a_3^2, \label{Eq:zl2}
\end{align}
where,
\begin{align*}
a_1 &= \mathcal{Z}_{L=1},~and \\
a_{2(3)} &= M_{ab}^2 - 2M_{D(a)}^2 - 2M_{c(b)}^2 + \frac{(M_{D(a)}^2 - M_{c(b)}^2)^2}{M_{ab}^2},
\end{align*}
and $M_{ij}$ denotes invariant mass of the two-particle system $ij$.

\clearpage

\section{Fits performed on individual tag-mode event samples separately}
\label{appensec:septagfits}
U-spin breaking parameter results obtained upon fitting each of the three tag-mode events separately to the corresponding amplitude function are presented along with their residuals, $(x_{\rm tot} - x_{\rm sep})/(\sigma(x_{\rm tot}) + \sigma(x_{\rm sep}))$, where $x_{\rm tot}$ are the fit parameters from the main fit using all the tag modes and $x_{\rm sep}$ are from the individual fits on the three tag-mode events.

\begin{table}[h]
\centering
\begin{tabular}{|c|c|c|c|c|}
\hline
Resonance & $r$ & Residual & $\delta$ (deg.) & Residual \Tstrut \\
\hline
$\rho(770)$     & $2.23 \pm 0.63$ & 0.33  & $-106.20 \pm 11.65$ & 0.89 \Tstrut \\
$\omega(782)$   & $7.93 \pm 3.59$ & 0.41  & $-19.26  \pm 25.86$ & 0.65 \Tstrut \\
$f_{2}(1270)$   & $4.44 \pm 2.07$ & 0.23  & $-86.57  \pm 35.95$ & 0.57 \Tstrut \\
$\rho(1450)$    & $17.77\pm 5.99$ & 0.63 & $~50.32  \pm 23.41$ & 0.74 \Tstrut \\
$\pi\pi$ S-wave & $0.29 \pm 0.67$  & 0.09 & $-188.8   \pm 55.12$ & 0.75 \Tstrut \\
\hline
\end{tabular}
\caption{U-spin breaking parameter values from individual fit on $K\pi$ tag mode events.}
\end{table}

\begin{table}[h]
\centering
\begin{tabular}{|c|c|c|c|c|}
\hline
Resonance & $r$ & Residual & $\delta$ (deg.) & Residual \Tstrut \\
\hline
$\rho(770)$     & $2.49\pm 0.53$ & 0.70 & $-79.94  \pm 8.36$  & 0.75 \Tstrut \\
$\omega(782)$   & $5.31\pm 1.27$ & 0.40 & $-13.61  \pm 14.93$ & 0.72 \Tstrut \\
$f_{2}(1270)$   & $2.30\pm 1.95$ & 0.51 & $-44.04  \pm 21.88$ & 0.32 \Tstrut \\
$\rho(1450)$    & $7.83\pm 4.07$ & 0.61 & $~118.5  \pm 39.04$ & 0.75 \Tstrut \\
$\pi\pi$ S-wave & $0.47\pm 0.56$ & 0.13 & $-111.8  \pm 58.83$ & 0.70 \Tstrut \\
\hline
\end{tabular}
\caption{U-spin breaking parameter values from individual fit on $K\pi\pi\pi$ tag mode events.}
\end{table}

\begin{table}[h]
\centering
\begin{tabular}{|c|c|c|c|c|}
\hline
Resonance & $r$ & Residual & $\delta$ (deg.) & Residual \Tstrut \\
\hline
$\rho(770)$     & $1.52\pm 0.40$ & 0.61 & $-82.57 \pm 10.58$ & 0.49 \Tstrut \\
$\omega(782)$   & $6.86\pm 0.97$ & 0.42 & $~14.38 \pm 7.61$  & 0.83 \Tstrut \\
$f_{2}(1270)$   & $4.54\pm 1.22$ & 0.37 & $-42.07 \pm 16.59$ & 0.43 \Tstrut \\
$\rho(1450)$    & $9.45\pm 2.74$ & 0.47 & $~78.05 \pm 19.45$ & 0.01 \Tstrut \\
$\pi\pi$ S-wave & $0.72\pm 0.38$ & 0.59 & $-206.90\pm 35.53$ & 0.83 \Tstrut \\
\hline
\end{tabular}
\caption{U-spin breaking parameter values from individual fit on $K\pi\pi^0$ tag mode events.}
\end{table}

\section{Standalone \texorpdfstring{$K_{\rm L}^0\pi^+\pi^-$}{klpipi} fit}
\label{appensec:klonly}
An amplitude fit is performed on the $K_{\rm L}^0\pi^+\pi^-$ data events alone without using any constraint related to the $K_{\rm S}^0\pi^+\pi^-$ amplitude. To avoid redundancy, no U-spin breaking parameters ($r_k, \delta_k$) are multiplied to the CP eigenstate amplitudes as separate fit parameters, except for the reference $\rho(770)$ and the $\pi\pi$ S-wave since they are as usual kept fixed in the fit. The fixed and floated U-spin breaking parameters are given in table~\ref{tab:rdelta_vals_klonly} and the resultant fit fractions and their comparison with the ones from the main fit are given in table~\ref{tab:k0lonly_ffr}. 

\begin{table}[h]
\centering
\begin{tabular}{|c|c|c|}
\hline
Resonance & Simultaneous fit results ($r, \delta$) & $K_{\rm L}^0\pi^+\pi^-$-only fit results ($r, \delta$)  \Tstrut \\
\hline
$\rho(770)$     & $1.93\pm 0.27,$   $-90.61 \pm 5.83$  & 	$1.93\pm1.10,~-99.9\pm 22.9$			\Tstrut \\
$\omega(782)$   & $6.13\pm 0.75,$   $~2.24  \pm 6.99$ 	& $1.0, 0.0$ (fixed)			\Tstrut \\
$f_{2}(1270)$   & $3.75\pm 0.90,$   $-56.52 \pm 16.84$ 	& $1.0, 0.0$ (fixed)			\Tstrut \\
$\rho(1450)$    & $12.12\pm 2.92,$  $~78.37 \pm 14.36$  & $1.0, 0.0$ (fixed)			\Tstrut \\
$\pi\pi$ S-wave & $0.37\pm 0.21,$   $-164.36 \pm 15.69$ & $0.91\pm 0.99,~-127.6\pm 61.9$ 			\Tstrut \\
\hline
\end{tabular}
\caption{U-spin breaking parameter values from a standalone fit.}
\label{tab:rdelta_vals_klonly}
\end{table}

\begin{table}[h]
\centering
\begin{tabular}{|c|c|c|}
\hline
Resonance & $K_{\rm L}^0\pi^+\pi^-$ FF($\%$) (Simultaneous fit) & $K_{\rm L}^0\pi^+\pi^-$ FF($\%$) (Standalone $K_{\rm L}^0\pi^+\pi^-$) \Tstrut \Bstrut \\
\hline
$\rho(770)$           & $18.16^{+0.53}_{-0.45}$ 		& $18.94\pm 1.20$ \Tstrut \\
$\omega(782)$         & $0.06_{-0.02}^{+0.03}$         & $0.06 \pm 0.03$ \Tstrut \\
$f_{2}(1270)$         & $0.40 \pm 0.08$   				& $0.36 \pm 0.08$ \Tstrut \\
$\rho(1450)$          & $0.42 \pm 0.08$         		& $0.43 \pm 0.10$ \Tstrut \\
$K^{*}(892)^{-}$      & $56.98^{+0.58}_{-0.56}$        & $57.1 \pm 1.65$ \Tstrut \\
$K_{2}^{*}(1430)^{-}$ & $1.64^{+0.10}_{-0.09}$         & $1.58 \pm 0.15$ \Tstrut \\
$K^{*}(1680)^{-}$     & $0.25^{+0.06}_{-0.05}$  		& $0.22 \pm 0.11$ \Tstrut \\
$K^{*}(1410)^{-}$     & $0.19 \pm 0.06$  				& $0.11 \pm 0.06$ \Tstrut \\
$K^{*}(892)^{+}$      & $0.45 \pm 0.05$  				& $0.37 \pm 0.06$ \Tstrut \\
$K_{2}^{*}(1430)^{+}$ & $0.05 \pm 0.02$  				& $0.02 \pm 0.02$ \Tstrut \\
$K^{*}(1410)^{+}$     & $0.04 \pm 0.02$         		& $0.02 \pm 0.02$ \Tstrut \\ 
$K_{0}^{*}(1430)^{-}$ & $6.84^{+0.24}_{-0.25}$         & $5.80 \pm 0.38$ \Tstrut \Bstrut \\
\hline
$\pi\pi$ S-wave       & $10.12_{-0.33}^{+0.32}$ 		& $10.39 \pm 1.72$ \Tstrut \Bstrut \\
\hline \hline
Total FF              & $95.59^{+2.16}_{-2.07}$ 		& $95.40 \pm 5.58$ \Tstrut \Bstrut \\
\hline
\end{tabular}
\caption{$D^0 \to K_{\rm L}^0\pi^+\pi^-$ fit fractions from a standalone fit.}
\label{tab:k0lonly_ffr}
\end{table}

\clearpage

\section{Systematic source-wise break up}
\label{appensec:systbreakup}

\begin{table}[h!]
\centering
\begin{tabular}{c|c}
\hline
                & Source   \Tstrut \Bstrut \\
\hline \hline
I  & Acceptance  \Tstrut \Bstrut \\
\hline
II & Resonance masses and widths, fixed LASS parameters, \Tstrut \\
   & and tag-side strong-phase parameters \Bstrut  \\ 
\hline
III   & Radii of resonance and $D$ meson  \Tstrut \Bstrut \\
\hline
IV    & Peaking background fraction    \Tstrut \Bstrut \\
\hline
V     & Non-peaking background shape  \Tstrut \Bstrut \\
\hline
VI    & K-matrix coupling ($\beta$) and production parameters ($f_{prod}$) \Tstrut \Bstrut \\
 \hline
\end{tabular}
\caption{Nomenclature for systematic sources.}
\end{table}

\begin{table}[h]
\centering
\small
\begin{tabular}{|c|c|cccccc|}
\hline
Resonance & $r$ & I & II & III & IV & V & VI \Tstrut \\
\hline
$\rho(770)$     & $1.93\pm0.27$  & 0.15 & 0.98  & 0.07 & 0.004 & 0.04 & 1.18  \Tstrut \\
$\omega(782)$   & $6.13\pm0.75$  & 0.13 & 0.57  & 0.15 & 0.005 & 0.01 & 0.36  \Tstrut \\
$f_{2}(1270)$   & $3.75\pm0.90$  & 0.22 & 0.39  & 0.07 & 0.002 & 0.01 & 0.78  \Tstrut \\
$\rho(1450)$    & $12.12\pm2.92$ & 0.11 & 0.52  & 0.12 & 0.001 & 0.004 & 0.34  \Tstrut \\
$\pi\pi$ S-wave & $0.37\pm0.21$  & 0.19 & 1.02  & 0.04 & 0.004 & 0.00 & 1.43  \Tstrut \\
\hline
\end{tabular}
\caption{Systematic uncertainties in units of statistical uncertainties on U-spin breaking parameters $r$.}
\label{tab:sys_uspin_r}
\end{table}

\begin{table}[h]
\centering
\small
\begin{tabular}{|c|c|cccccc|}
\hline
Resonance & $\delta$ (deg.) & I & II & III & IV & V & VI \Tstrut \\
\hline
$\rho(770)$     & $-90.61  \pm 5.83$  & 0.63 & 0.70 & 0.23 & 0.003 & 0.10 & 0.87 \Tstrut \\
$\omega(782)$   & $~2.24   \pm 6.99$  & 0.01 & 0.50 & 0.04 & 0.003 & 0.05 & 0.47 \Tstrut \\
$f_{2}(1270)$   & $-56.52  \pm 16.84$ & 0.05 & 0.59 & 0.21 & 0.001 & 0.04 & 0.44 \Tstrut \\
$\rho(1450)$    & $~78.37  \pm 14.36$ & 0.18 & 0.92 & 0.13 & 0.001 & 0.04 & 0.53 \Tstrut \\
$\pi\pi$ S-wave & $-164.36 \pm 15.69$ & 0.43 & 0.62 & 0.05 & 0.003 & 0.03 & 0.40 \Tstrut \\
\hline
\end{tabular}
\caption{Systematic uncertainties in units of statistical uncertainties on U-spin breaking parameters $\delta$.}
\label{tab:sys_uspin_delta}
\end{table}

\begin{table}[h]
\centering
\begin{tabular}{|c|c|cccccc|}
\hline
Resonance & $K_{\rm S}^0\pi^+\pi^-$ FF($\%$) & I & II & III & IV & V & VI \Tstrut \Bstrut \\
\hline
$\rho(770)$            & $18.90\pm 0.42$ & 1.21 & 0.54 & 0.43 & 0.0003 & 0.07 & 2.80 \Tstrut \\
$\omega(782)$          & $0.54\pm 0.09$ & 0.11 & 0.38 & 0.34 & 0.001 & 0.00 & 0.67 \Tstrut \\
$f_{2}(1270)$          & $0.61_{-0.11}^{+0.13}$ & 0.40 & 0.32 & 0.48 & 0.004 & 0.00 & 1.20 \Tstrut \\
$\rho(1450)$           & $0.21 \pm 0.10$ & 0.16 & 0.64 & 0.72 & 0.001 & 0.08 & 2.4 \Tstrut \\
$K^{*}(892)^{-}$       & $62.18_{-0.59}^{0.55}$ & 0.37 & 0.02 & 0.51 & 0.001 & 0.04 & 3.58 \Tstrut \\
$K_{2}^{*}(1430)^{-}$  & $1.79 \pm 0.09$ & 0.20 & 0.01 & 0.30 & 0.0001 & 0.80 & 3.90 \Tstrut \\
$K^{*}(1680)^{-}$      & $0.27\pm 0.06$ & 2.33 & 2.86 & 2.17 & 0.0003 & 0.17 & 3.00 \Tstrut \\
$K^{*}(1410)^{-}$      & $0.21 \pm 0.06$ & 0.33 & 1.86 & 0.17 & 0.001 & 0.17 & 0.67 \Tstrut \\
$K^{*}(892)^{+}$       & $0.49\pm 0.05$ & 1.00 & 2.00 & 0.50 & 0.005 & 0.50 & 3.00 \Tstrut \\
$K_{2}^{*}(1430)^{+}$  & $0.05 \pm 0.02$ & 0.00 & 0.22 & 0.00 & 0.006 & 0.50 & 1.00 \Tstrut \\
$K^{*}(1410)^{+}$      & $0.05 \pm 0.03$ & 0.00 & 0.18 & 0.00 & 0.002 & 0.33 & 0.33 \Tstrut \\ 
$K_{0}^{*}(1430)^{-}$  & $7.47 \pm 0.26$ & 0.3 & 0.86 & 0.46 & 0.0003 & 0.04 & 4.29 \Tstrut \Bstrut \\
\hline

$\pi\pi$ S-wave        & $10.24\pm 0.23$ & 0.58 & 0.74 & 0.83 & 0.0004 & 0.17 & 4.71 \Tstrut \Bstrut \\
\hline
\end{tabular}
\caption{Systematic uncertainties in units of statistical uncertainties on $K_{\rm S}^0\pi^+\pi^-$ fit fractions.}
\label{tab:sys_ks_ffr}
\end{table}

\begin{table}[h]
\centering
\begin{tabular}{|c|c|cccccc|}
\hline
Resonance & $K_{\rm L}^0\pi^+\pi^-$ FF($\%$) & I & II & III & IV & V & VI \Tstrut \Bstrut \\
\hline
$\rho(770)$           & $18.16^{+0.53}_{-0.45}$ & 0.38 & 1.02 & 0.66 & 0.0004 & 0.51 & 2.53 \Tstrut \\
$\omega(782)$         & $0.06 \pm 0.025$ & 0.00 & 0.40  & 0.92 & 0.002 & 0.00 & 0.40 \Tstrut \\
$f_{2}(1270)$         & $0.40 \pm 0.08$ & 0.53 & 1.16 & 0.40 & 0.006 & 0.27 & 2.27 \Tstrut \\
$\rho(1450)$          & $0.42\pm 0.08$ & 0.50 & 1.93 & 1.00 & 0.004 & 0.12 & 3.12  \Tstrut \\
$K^{*}(892)^{-}$      & $56.98^{+0.58}_{-0.56}$ & 0.82 & 1.68 & 0.44 & 0.003 & 0.93 & 1.57 \Tstrut \\
$K_{2}^{*}(1430)^{-}$ & $1.64^{+0.10}_{-0.09}$ & 0.00 & 0.67 & 0.21 & 0.0005 & 0.53 & 3.68 \Tstrut \\
$K^{*}(1680)^{-}$     & $0.25^{+0.06}_{-0.05}$ & 2.36 & 2.67 & 2.18 & 0.0001 & 1.82 & 3.27 \Tstrut \\
$K^{*}(1410)^{-}$     & $0.19\pm 0.06$ & 0.33  & 1.93 & 0.17 & 0.001 & 0.17 & 5.00 \Tstrut \\
$K^{*}(892)^{+}$      & $0.45\pm 0.05$ & 0.40 & 0.80 & 0.20 & 0.005 & 0.20 & 1.20 \Tstrut \\
$K_{2}^{*}(1430)^{+}$ & $0.05\pm 0.02$ & 0.50  & 0.21 & 0.00 & 0.005 & 0.50 & 1.00 \Tstrut \\
$K^{*}(1410)^{+}$     & $0.04 \pm 0.02$ & 0.00 & 0.51  & 0.00 & 0.004 & 0.50 & 0.50 \Tstrut \\ 
$K_{0}^{*}(1430)^{-}$ & $6.84^{+0.24}_{-0.25}$ & 0.65 & 1.37 & 0.49 & 0.001 & 0.78 & 4.20 \Tstrut \Bstrut \\
\hline
$\pi\pi$ S-wave       & $10.12^{+0.32}_{-0.33}$ & 0.37 & 0.68 & 0.25 & 0.004 & 0.18 & 1.46 \Tstrut \Bstrut \\
\hline 
\end{tabular}
\caption{Systematic uncertainties in units of statistical uncertainties on $K_{\rm L}^0\pi^+\pi^-$ fit fractions.}
\label{tab:sys_kl_ffr}
\end{table}

\begin{table}[h]
\centering
\begin{tabular}{|c|c|cccccc|}
\hline
Bin & $c_i$ &  I & II & III & IV & V & VI  \Tstrut \\
\hline
1 &  $~0.695\pm0.007$ & 0.00 & 1.30 & 0.62 & 0.003 & 0.18 & 1.22  \Tstrut \\
2 &  $~0.614\pm0.010$ & 0.39 & 1.32 & 0.57 & 0.002 & 0.65 & 1.28  \Tstrut \\
3 &  $~0.038\pm0.027$ & 0.46 & 2.66 & 0.37 & 0.001 & 0.46 & 2.24  \Tstrut \\
4 &  $-0.542\pm0.027$ & 0.39 & 0.97 & 0.24 & 0.001 & 0.41 & 2.70 \Tstrut \\
5 &  $-0.947\pm0.004$ & 0.11 & 1.85 & 0.51 & 0.012 & 0.06 & 2.92  \Tstrut \\
6 &  $-0.494\pm0.020$ & 1.21 & 0.98 & 0.21 & 0.0005 & 0.007 & 1.98  \Tstrut \\
7 &  $~0.114\pm0.026$ & 0.67 & 1.52 & 0.40 & 0.002 & 0.17 & 1.90  \Tstrut \\
8 &  $~0.453\pm0.015$ & 0.48 & 1.87 & 0.41 & 0.003 & 0.10 & 2.18  \Tstrut \\
\hline
\end{tabular}
\caption{Systematic uncertainties in units of statistical uncertainties on $c_i$.}
\label{tab:sys_ci}
\end{table}

\begin{table}[h]
\centering
\begin{tabular}{|c|c|cccccc|}
\hline
Bin & $s_i$ & I & II & III & IV & V & VI \Tstrut \\
\hline
1 &	$~0.021\pm0.014$ & 0.82 & 0.75 & 0.64 & 0.001 & 0.15 & 1.64 \Tstrut \\
2 & $~0.408\pm0.013$ & 0.75 & 1.09 & 0.84 & 0.002 & 0.15 & 1.95 \Tstrut \\
3 & $~0.831\pm0.006$ & 1.02 & 1.82 & 0.08 & 0.002 & 0.37 & 2.55 \Tstrut \\
4 &	$~0.719\pm0.020$ & 0.27 & 0.57 & 0.37 & 0.001 & 0.40 & 2.56 \Tstrut \\
5 & $-0.052\pm0.020$ & 0.06 & 0.85 & 0.24 & 0.001 & 0.03 & 2.08 \Tstrut \\
6 & $-0.644\pm0.012$ & 0.28 & 0.52 & 0.51 & 0.002 & 0.13 & 2.29 \Tstrut \\
7 & $-0.805\pm0.014$ & 1.51 & 0.97 & 0.73 & 0.001 & 0.17 & 1.14 \Tstrut \\
8 & $-0.441\pm0.021$ & 1.02 & 0.37 & 0.72 & 0.001 & 0.04 & 1.42 \Tstrut \\
\hline
\end{tabular}
\caption{Systematic uncertainties in units of statistical uncertainties on $s_i$.}
\label{tab:sys_si}
\end{table}

\begin{table}[h]
\centering
\begin{tabular}{|c|c|cccccc|}
\hline
Bin  & $c_i'$ & I & II & III & IV & V & VI \Tstrut \\
\hline
1 &	$~0.797\pm0.006$ & 0.69 & 1.39 & 0.52 & 0.002 & 1.14 & 1.23  \Tstrut \\
2 & $~0.812\pm0.007$ & 0.23 & 1.22 & 0.42 & 0.005 & 0.24 & 1.51  \Tstrut \\
3 & $~0.292\pm0.025$ & 0.46 & 0.84 & 0.49 & 0.002 & 0.30 & 1.19  \Tstrut \\
4 & $-0.392\pm0.027$ & 0.34 & 0.70 & 0.36 & 0.001 & 0.26 & 1.32  \Tstrut \\
5 & $-0.923\pm0.005$ & 0.13 & 2.61 & 0.67 & 0.002 & 2.49 & 3.72  \Tstrut \\
6 & $-0.300\pm0.020$ & 1.21 & 1.01 & 0.36 & 0.002 & 0.77 & 2.39  \Tstrut \\
7 & $~0.541\pm0.020$ & 2.50 & 2.77 & 0.32 & 0.003 & 0.11 & 2.66  \Tstrut \\
8 & $~0.726\pm0.010$ & 2.24 & 2.14 & 0.50 & 0.003 & 0.26 & 2.49  \Tstrut \\
\hline
\end{tabular}
\caption{Systematic uncertainties in units of statistical uncertainties on $c_i'$.}
\label{tab:sys_cip}
\end{table}

\begin{table}[h]
\centering
\begin{tabular}{|c|c|cccccc|}
\hline
Bin  & $s_i'$ & I & II & III & IV & V & VI \Tstrut \\
\hline
1 & $~0.044\pm0.014$ & 1.20 & 0.90 & 0.40 & 0.002 & 0.81 & 1.55 \Tstrut \\
2 & $~0.327\pm0.014$ & 0.60 & 1.58 & 0.52 & 0.004 & 0.19 & 1.72 \Tstrut \\
3 & $~0.806\pm0.010$ & 0.13 & 1.88 & 0.36 & 0.004 & 0.08 & 1.93 \Tstrut \\
4 & $~0.811\pm0.015$ & 0.48 & 0.56 & 0.43 & 0.001 & 0.85 & 2.02 \Tstrut \\
5 & $-0.045\pm0.040$ & 0.03 & 0.74 & 0.21 & 0.001 & 0.27 & 1.63 \Tstrut \\
6 & $-0.641\pm0.013$ & 0.26 & 0.84 & 0.49 & 0.002 & 0.18 & 1.85 \Tstrut \\
7 & $-0.663\pm0.028$ & 0.66 & 1.58 & 0.64 & 0.002 & 0.83 & 1.88 \Tstrut \\
8 & $-0.327\pm0.020$ & 1.21 & 0.38 & 0.54 & 0.002 & 0.27 & 1.62 \Tstrut \\
\hline
\end{tabular}
\caption{Systematic uncertainties in units of statistical uncertainties on $s_i'$.}
\label{tab:sys_sip}
\end{table}

\clearpage

\section{Predicted and assumed \texorpdfstring{$\Delta c_i, \Delta s_i$}{DciDsi}}
\label{appen:DciDsi}

\begin{table}[h]
    \centering
    \begin{tabular}{c|c|c|c|c}
    \hline \hline
        Bin & $\Delta c_{i}$ $\pm$ $\delta \Delta c_i$ & $\Delta s_i$ $\pm$ $\delta \Delta s_i$ & $\Delta c_{i}$ $\pm$ $\delta \Delta c_i$ & $\Delta s_i$ $\pm$ $\delta \Delta s_i$  \Tstrut \\
            & [predicted] & [predicted] & [assumed] & [assumed] \Bstrut \\
        \hline \hline
        1 & $0.12\pm0.02$ & $~0.03\pm0.03$ & $0.12\pm0.02$ & $-0.02\pm0.02$ \Tstrut \\    
        2 & $0.20\pm0.02$ & $-0.08\pm0.04$ & $0.19\pm0.02$ & $-0.06\pm0.07$ 	\\
        3 & $0.26\pm0.07$ & $-0.02\pm0.04$ & $0.28\pm0.10$ & $~0.01\pm0.02$	\\
        4 & $0.15\pm0.10$ & $~0.09\pm0.05$ & $0.15\pm0.14$ & $~0.11\pm0.09$	\\
        5 & $0.02\pm0.01$ & $~0.00\pm0.03$ & $0.04\pm0.05$ & $~0.02\pm0.06$	\\
        6 & $0.19\pm0.03$ & $~0.00\pm0.03$ & $0.27\pm0.16$ & $-0.02\pm0.05$	\\
        7 & $0.42\pm0.05$ & $~0.14\pm0.04$ & $0.48\pm0.07$ & $~0.15\pm0.07$	\\
        8 & $0.27\pm0.03$ & $~0.11\pm0.03$ & $0.29\pm0.04$ & $~0.07\pm0.04$	\\
    \hline
    \end{tabular}
    \caption{Model-predicted and previously assumed strong-phase parameter differences ($\Delta c_i, \Delta s_i$). Uncertainties on the predicted values include both statistical and systematic.}
    \label{tab:DciDsi}
\end{table}

\clearpage

\textbf{Acknowledgement}

The BESIII collaboration thanks the staff of BEPCII and the IHEP computing center for their strong support. This work is supported in part by National Key R\&D Program of China under Contracts Nos. 2020YFA0406300, 2020YFA0406400; National Natural Science Foundation of China (NSFC) under Contracts Nos. 11635010, 11735014, 11835012, 11935015, 11935016, 11935018, 11961141012, 12022510, 12025502, 12035009, 12035013, 12061131003, 12192260, 12192261, 12192262, 12192263, 12192264, 12192265; the Chinese Academy of Sciences (CAS) Large-Scale Scientific Facility Program; the CAS Center for Excellence in Particle Physics (CCEPP); Joint Large-Scale Scientific Facility Funds of the NSFC and CAS under Contract No. U1832207; CAS Key Research Program of Frontier Sciences under Contracts Nos. QYZDJ-SSW-SLH003, QYZDJ-SSW-SLH040; 100 Talents Program of CAS; The Institute of Nuclear and Particle Physics (INPAC) and Shanghai Key Laboratory for Particle Physics and Cosmology; ERC under Contract No. 758462; European Union's Horizon 2020 research and innovation programme under Marie Sklodowska-Curie grant agreement under Contract No. 894790; German Research Foundation DFG under Contracts Nos. 443159800, 455635585, Collaborative Research Center CRC 1044, FOR5327, GRK 2149; Istituto Nazionale di Fisica Nucleare, Italy; Ministry of Development of Turkey under Contract No. DPT2006K-120470; National Science and Technology fund; National Science Research and Innovation Fund (NSRF) via the Program Management Unit for Human Resources \& Institutional Development, Research and Innovation under Contract No. B16F640076; Olle Engkvist Foundation under Contract No. 200-0605; STFC (United Kingdom); Suranaree University of Technology (SUT), Thailand Science Research and Innovation (TSRI), and National Science Research and Innovation Fund (NSRF) under Contract No. 160355; The Royal Society, UK under Contracts Nos. DH140054, DH160214; The Swedish Research Council; U. S. Department of Energy under Contract No. DE-FG02-05ER41374


\end{document}